\documentclass[12pt,english]{article}
\usepackage[table]{xcolor}
\usepackage[T1]{fontenc}
\usepackage[utf8]{inputenc}
\usepackage[pdftex]{geometry}
\usepackage{bm}
\usepackage{amsmath}
\usepackage{amssymb}
\usepackage[pdftex]{graphicx}
\usepackage{setspace}
\usepackage{ytableau}

\makeatletter
\numberwithin{equation}{section}
\newcommand{\lyxaddress}[1]{
	\par {\raggedright #1
	\vspace{1.4em}
	\noindent\par}
}

\allowdisplaybreaks[1]

\usepackage{braketmod}

\usepackage{color}
\usepackage[pdftex]{hyperref}
\hypersetup{
colorlinks,
linktoc=page,
citecolor=blue,
linkcolor=blue,
urlcolor=blue
}

\date{}

\makeatother

\usepackage{babel}
\usepackage[style=phys,biblabel=brackets, chaptertitle=false, pageranges=false, eprint=true]{biblatex}
\begin{document}

\ytableausetup{boxsize=1.5mm}

\noindent\begin{minipage}[t]{1\columnwidth}%
\title{\textbf{Exact large $N$ expansion of mass deformed\\
ABJM theory on squashed sphere}}
\author{Naotaka Kubo\,\footnotemark,\quad{}Tomoki Nosaka\,\footnotemark\quad{}and\quad{}Yi Pang\,\footnotemark}
\maketitle

\lyxaddress{\begin{center}
\vspace{-18bp}
$^{*}$ $^{\ddagger}$\,\textit{Center for Joint Quantum Studies and Department of Physics, School of Science,}\\
\textit{Tianjin University, 135 Yaguan Road, Tianjin 300350, China}\\
\vspace{10bp}
$^{\dagger}$\,\textit{Shanghai Institute for Mathematics and Interdisciplinary Sciences,}\\
\textit{Block A, International Innovation Plaza, No.~657 Songhu Road, Yangpu District, Shanghai, China}\\
\vspace{10bp}
$^{\ddagger}$\,\textit{Peng Huanwu Center for Fundamental Theory, Hefei, Anhui 230026, China}
\vspace{-10bp}
\par\end{center}}
\begin{abstract}
In this paper we study the partition function of the mass deformed ABJM theory on a squashed three sphere.
In particular, we focus on the case with the Chern-Simons levels being $\pm 1$ and apply a duality between this theory and the $\mathcal{N}=4$ $\mathrm{U}\left(N\right)$ super Yang-Mills theory with an adjoint hypermultiplet and a fundamental hypermultiplet.
For a special mass parameter depending on the squashing parameter, we find that the partition function can be written as that of an ideal Fermi gas with a non-trivial density matrix.
By studying this density matrix, we analytically derive the all order perturbative expansion of the partition function in $1/N$, which turns out to take the form of the Airy function.
Our results not only align with previous findings and conjectures but also lead to a new formula for the overall constant factor of the partition function.
We also study the exact values of the partition function for small but finite values of $N$.
\end{abstract}
\end{minipage}

\renewcommand{\thefootnote}{\fnsymbol{footnote}}
\footnotetext[1]{\textsf{naotaka.kubo@yukawa.kyoto-u.ac.jp}}
\footnotetext[2]{\textsf{nosaka@yukawa.kyoto-u.ac.jp}}
\footnotetext[3]{\textsf{pangyi1@tju.edu.cn}}
\renewcommand{\thefootnote}{\arabic{footnote}}\medskip{}
\thispagestyle{empty}

\newpage{}

\setcounter{page}{1}

\global\long\def\bra#1{\Bra{#1}}%

\global\long\def\bbra#1{\Bbra{#1}}%

\global\long\def\ket#1{\Ket{#1}}%

\global\long\def\kket#1{\Kket{#1}}%

\global\long\def\braket#1{\Braket{#1}}%

\global\long\def\bbraket#1{\Bbraket{#1}}%

\global\long\def\brakket#1{\Brakket{#1}}%

\global\long\def\bbrakket#1{\Bbrakket{#1}}%

\global\long\def\cm{A^{\mathrm{c}}}%

\tableofcontents{}

\section{Introduction}

The AdS/CFT correspondence \cite{Maldacena:1997re} is a powerful tool for investigating quantum gravity.
This is an idea that quantum gravity can be analyzed via a conformal field theory living on the boundary of the anti-de Sitter space.
This idea is successfully realized by a stack of $N$ of D3-branes, whose two aspects from open/closed strings lead to a duality between the 4d $\mathcal{N}=4$ $\text{U}\left(N\right)$ super Yang-Mills theory and type IIB string theory on $\text{AdS}_5\times S^5$.
Another important duality comes from $N$ M2-branes, which leads to a duality between the $\text{U}\left(N\right)_k\times \text{U}\left(N\right)_{-k}$
ABJM theory \cite{Hosomichi:2008jd,Aharony:2008ug} with finite $k$ and M-theory on $\text{AdS}_{4}\times S^{7}/\mathbb{Z}_{k}$.
Here ABJM theory is a 3d $\mathcal{N}=6$ superconformal Chern-Simons theory which consists of a pair of $\mathcal{N}=2$ $\text{U}(N)$ Chern-Simons vector multiplets with the Chern-Simons levels $\pm k$ and four $\mathcal{N}=2$ chiral multiplets in the bi-fundamental representations of $\mathrm{U}\left(N\right)_k\times\mathrm{U}\left(N\right)_{-k}$.

In general, the leading term of an observable in the large $N$ limit is captured in the AdS side by the classical supergravity.
On the other hand, computing the $1/N$ corrections is very important for investigating quantum gravity effects, although their evaluation in a strongly coupled field theory is more challenging.
So far, the large $N$ behavior has been explored in great detail for the ABJM theory placed on the three sphere $S^3$.
An important breakthrough was the discovery of the supersymmetric localization technique \cite{Pestun:2007rz,Kapustin:2009kz,Jafferis:2010un,Hama:2010av,Hama:2011ea,Imamura:2011wg}, which reduces the path integral computing the sphere partition function to a matrix integral, which we shall call the ABJM matrix model.
The partition function admits two types of deformation which are compatible with the localization technique: the deformation of the round three sphere $S^3=S^3_{b=1}$ to a squashed three sphere $S^3_b$ ($b>1$) preserving $\text{U}(1)\times \text{U}(1)$ isometry, and the three-parameter real mass deformations $m_1,m_2,m_3$ corresponding to the Cartans of $\text{SU}(2)\times \text{SU}(2)\times \text{U}(1) \subset \text{SO}(6)$ global symmetry \cite{Aharony:1997bx}.
The deformed partition function gives a generating function of various observables, and can also be used to compute various quantities such as the supersymmetric R\'enyi entropy \cite{Nishioka:2013haa}, the two-point function of the stress energy tensor \cite{Closset:2012ru,Nishioka:2013gza} and the OPE coefficients of the stress tensor multiplet \cite{Agmon:2017xes,Binder:2018yvd,Binder:2019mpb,Chester:2020jay,Chester:2021gdw}.

The large $N$ behavior of the ABJM matrix model was first investigated in the 't Hooft limit ($N,k\rightarrow\infty$ with $N/k$ fixed) without deformations, namely $b=1$ and $m_I=0$ \cite{Drukker:2010nc}, where the $N^{3/2}$ scaling of the free energy, which is characteristic of the M2-branes \cite{Klebanov:1996un}, was reproduced including the coefficient obtained from the supergravity on $\text{AdS}_{4}$.
Surprisingly, it was further found that the all order $1/N$ perturbative expansion can be resummed by an Airy function \cite{Fuji:2011km}.

Notice that in the 't Hooft limit the gravity side reduces to the type IIA string theory on $\text{AdS}_4\times \mathbb{CP}^3$, while our main interest is the M-theory region, which corresponds to the large $N$ limit with $k$ fixed.
In this regard, another breakthrough was the Fermi gas formalism \cite{Marino:2011eh} which rewrites the ABJM matrix model into a partition function of an ideal Fermi gas with non-trivial density matrix.
In this formalism, the large $N$ limit corresponds to the thermodynamic limit, while the Chern-Simons level $k$ plays the role of the Planck’s constant.
Hence the Fermi gas formalism allows us to study the M-theory region systematically via ordinary techniques of quantum mechanics such as the WKB expansion.
The Fermi gas formalism was also utilized to reveal the M-theoretic large $N$ expansion of many different Chern-Simons matter theories which correspond to M2-branes placed on different backgrounds,
such as the ABJ theory \cite{Hosomichi:2008jb,Aharony:2008gk} with non-uniform ranks \cite{Awata:2012jb,Honda:2013pea,Matsumoto:2013nya,Honda:2014npa} and/or the orthogonal/symplectic gauge groups \cite{Mezei:2013gqa,Moriyama:2015asx,Honda:2015rbb,Okuyama:2016xke,Kubo:2024zug}, circular quiver Chern-Simons theories \cite{Marino:2011eh,Drukker:2015awa} and the theories with affine D-type quiver diagrams \cite{Assel:2015hsa,Moriyama:2015jsa,Kubo:2024raz}.

As we have seen above, the Fermi gas formalism is an effective approach to study the M-theory region.
It would be natural to ask whether one can apply the Fermi gas formalism also for the ABJM matrix model with squashing/mass deformation.
Indeed, the Fermi gas formalism was successfully applied and the exact large $N$ expansion of the partition function was obtained up to the non-perturbative corrections when $b=1$ and only two of the three mass parameters are turned on, $m_3=0$ \cite{Nosaka:2015iiw}.
On the other hand, turning on the squashing parameter $b\neq1$ makes it difficult to apply the Fermi gas formalism, and hence only a few results are known about the large $N$ expansion.\footnote{
The coefficient of the leading term $N^{3/2}$ in the free energy $-\log Z$ can also be obtained by using the large $N$ saddle point approximation \cite{Herzog:2010hf}.
This method is applicable even for a generic value of the squashing parameter $b$ \cite{Imamura:2011wg,Shimizu:2018evl}.
}
For special value of $b^{2}=3$, the Fermi gas formalism can be applied and the exact large $N$ perturbative expansion was obtained \cite{Hatsuda:2016uqa}.
It is also known that the matrix model is invariant under the squashing when $m_3$ is adjusted as a function of $b$, for which the large $N$ expansion trivially coincides with that for $b=1$ with $m_3=0$ \cite{Chester:2021gdw} (see also \cite{Minahan:2021pfv}).

An interesting fact is that in all the known results discussed above which were obtained from the Fermi gas formalism, the all order $1/N$ perturbative expansion has the following form of the Airy function
\begin{equation}
Z^{\mathrm{ABJM}}\left(N\right)\approx
Z^{\mathrm{ABJM}\left(\mathrm{pert}\right)}\left(N\right)=e^{A}C^{-\frac{1}{3}}\mathrm{Ai}\left[C^{-\frac{1}{3}}\left(N-B\right)\right].\label{eq:AiryForm}
\end{equation}
(The first equality is up to the non-perturbative corrections in $1/N$ to the free energy $-\log Z^{\mathrm{ABJM}}(N)$.)
This function is characterized by only three parameters $A$, $B$ and $C$.
Hence, it would be natural to assume that the exact $1/N$ perturbative expansion of the ABJM matrix model with general parameters $\left(k,b,m_{1,2,3}\right)$ keeps the Airy function form \eqref{eq:AiryForm}.
Based on this assumption, together with the symmetry of the three mass parameters and the analysis in the supegravity including the higher-derivative corrections \cite{Bobev:2021oku}, the expressions for the two parameters $B$ and $C$ for generic values of $\left(k,b,m_{1,2,3}\right)$ were proposed as \eqref{eq:BC-ABJM} \cite{Bobev:2022jte,Hristov:2022lcw,Bobev:2022eus}, which are consistent with the previous results in such as \cite{Nosaka:2015iiw,Chester:2021gdw}.
We will review this proposal in section \ref{subsec:LargeN-Airy}.

Although the conjecture provides the exact $1/N$ perturbative expansion, there are still several unsatisfactory points.
First, the expression for $A$ has not been proposed for generic values of $\left(k,b,m_{1,2,3}\right)$ in a consistent manner with the known result for $b^2=3$ \cite{Hatsuda:2016uqa}.
Note that although $A$ gives only an $N$ independent overall factor of the partition function, it affects the observables obtained by taking derivatives of the partition function with respect to the deformation parameters $b$ and $m_{1,2,3}$.
The contribution from $A$ is indeed crucial, for example, in the study of the OPE coefficients \cite{Agmon:2017xes,Binder:2018yvd,Binder:2019mpb,Chester:2020jay,Chester:2021gdw}.
Second, the conjecture has not been proved yet.
In particular, there is no reason from the field theory side why the all of the coefficients of the $1/N$ expansion depend on $(k,b,m_I)$ only through the three parameters $A,B,C$ and are further fixed by the Airy function \eqref{eq:AiryForm}.\footnote{
See \cite{Dabholkar:2014wpa,Caputa:2018asc} for several attempts to understand the appearance of the Airy function from the gravity side.
}

In order to improve this situation, in this paper we rigorously obtain the all order $1/N$ perturbative expansion of the $S_{b}^{3}$ partition function of the ABJM theory with two free parameters including the squashing parameter $b$ by applying the Fermi gas formalism.
More concretely, we restrict the parameters to $(k,b,m_1,m_2,m_3)=(1,b,\zeta-m_{b},\zeta+m_{b},0)$ with two free parameters $\left(b,\zeta\right)$, where $m_{b}$ is defined in \eqref{eq:Mass-FGF}.
With these two free parameters we confirm that the exact $1/N$ perturbative expansion has the Airy form \eqref{eq:zADHMpert}, with $A,B,C$ given as \eqref{eq:ABC-FGF}.
Our results for $B$ and $C$ corroborate the previous conjecture \eqref{eq:BC-ABJM} for the above parameter region.
Furthermore, we obtain $A$, which is a new result and consistent with the known result for $(k,b,m_{1,2,3})=(1,\sqrt{3},0)$ \cite{Hatsuda:2016uqa}.

The reason why we restrict the Chern-Simons level to $k=1$ for obtaining our result is to use a duality between the $k=1$ ABJM theory and the $\mathcal{N}=4$ $\mathrm{U}\left(N\right)$ super Yang-Mills (SYM) theory with an adjoint hypermultiplet and a fundamental hypermultiplet.
This duality leads to an equality between the $k=1$ ABJM matrix model and the SYM matrix model which is also obtained by applying the supersymmetric localization to the $S_{b}^{3}$ partition function of the $\mathcal{N}=4$ SYM theory.
From the viewpoint of the matrix model, this equality is non-trivial, but conversely that means that one can deal with the SYM matrix model, which is relatively simple, instead of the relatively complicated ABJM matrix model.
We also give some pieces of evidence of the equality.
Then, by further restricting the mass parameters to $\left(m=m_{b},m_{3}=0\right)$, we successfully apply the Fermi gas formalism.

We also show that the SYM matrix model can be used for obtaining (closed form) exact values for odd $b^{2}$ as a function of $\left(m,\zeta\right)$ in a different way when $N=1,2,3$.
For small $N$ the SYM matrix model can be computed by directly evaluating the residues, and recently it was found that the residue can be classified by the combinatorics of Young diagram for $b=1$ \cite{Gaiotto:2019mmf}.
We find that a similar combinatorics works for odd $b^2$ at least up to $N=3$.

After evaluating the matrix model in two ways, we give some numerical checks for these discoveries.
We verify that these two results agree with each other.
Although we apply the Fermi gas formalism only for odd $b^{2}$ cases, the comparison with the exact values of the partition function for $N=1$ supports
that the Airy form we obtained \eqref{eq:zADHMpert} is correct for arbitrary $b>1$.
In addition, the exact values of the partition function for $N=1,2,3$ suggest that the large $N$ partition function obeys the Airy form even when $m\neq m_{b}$ with $C$ and $B$ given as \eqref{eq:BC-ADHM} and some undetermined constant $A$.

This paper is organized as follows.
In section \ref{sec:MM-LargeN}, we review the matrix model, the duality and the large $N$ conjecture.
In section \ref{sec:LargeN}, we apply the Fermi gas formalism to the SYM matrix model and obtain the exact large $N$ perturbative expansion.
In section \ref{sec:NumericComp}, we study the SYM matrix model with finite $N$.
In section \ref{subsec:NumCheck}, we perform some numerical tests for our results.
In section \ref{sec:Conclusion}, we summarize our work and give possible future directions.
In appendix \ref{sec:DSF}, we provide properties of the double sine function and a related function.
In section \ref{sec:Duality}, we check the duality between the $k=1$ ABJM theory and the $\mathcal{N}=4$ SYM theory and give parameter correspondence.

\section{Matrix model and large $N$\label{sec:MM-LargeN}}

As discussed in the introduction, the ABJM theory with the specific Chern-Simons level $k=1$ is dual to the $\mathcal{N}=4$ $\mathrm{U}\left(N\right)$ SYM theory with an adjoint hypermultiplet and a fundamental hypermultiplet.
Their partition functions on squashed three sphere $S_{b}^{3}$ reduce to matrix models by using the localization technique \cite{Kapustin:2009kz,Hama:2010av,Jafferis:2010un,Hama:2011ea}, and due to the duality they are expected to be equal.
The large $N$ perturbative expansion of the ABJM theory was conjectured in \cite{Bobev:2022eus}, and the prediction is interpreted as a prediction for the $\mathcal{N}=4$ SYM theory via the duality.
In this section we review these aspects.

\subsection{Matrix models\label{subsec:MatrixModel}}

In this section we review the matrix models of the ABJM theory and the $\mathcal{N}=4$ $\mathrm{U}\left(N\right)$ SYM theory  representing their partition functions on $S_{b}^{3}$.
We also review the duality between them.

In $\mathcal{N}=2$ language, the four chiral multiplets of the ABJM theory $Z^A,W_A$ ($A=1,2$) transform under the $\mathrm{SU}\left(2\right)\times\mathrm{SU}\left(2\right)\times\mathrm{U}\left(1\right)$ flavor symmetry, and thus the ABJM theory admits three mass deformations.
The matrix model for the ABJM theory on $S_{b}^{3}$ is \cite{Chester:2021gdw,Bobev:2022eus}
\begin{align}
 & Z_{k,b}^{\mathrm{ABJM}}\left(N;m_{1},m_{2},m_{3}\right)\nonumber \\
 & =\frac{1}{\left(N!\right)^{2}}\int_{\mathbb{R}}\prod_{i}^{N}\frac{d\mu_{i}}{2\pi}\frac{d\nu_{i}}{2\pi}e^{-\frac{ik}{4\pi}\sum_{i}^{N}\left(\mu_{i}^{2}-\nu_{i}^{2}\right)}
 \prod_{i<j}^{N}\prod_{\pm}\left\{\left(2\sinh\frac{b^{\pm 1}\left(\mu_{i}-\mu_{j}\right)}{2}\right)\left(2\sinh\frac{b^{\pm 1}\left(\nu_{i}-\nu_{j}\right)}{2}\right)\right\}\nonumber \\
 & \quad\quad\times\prod_{i,j}^{N}\left\{s_{b}\left(\frac{iQ}{4}+\left(\frac{\mu_{i}-\nu_{j}}{2\pi}+\frac{m_{2}+m_{1}-m_{3}}{2}\right)\right)s_{b}\left(\frac{iQ}{4}-\left(\frac{\mu_{i}-\nu_{j}}{2\pi}+\frac{m_{2}+m_{1}+m_{3}}{2}\right)\right)\right.\nonumber \\
 & \quad\quad\quad\times\left.s_{b}\left(\frac{iQ}{4}+\left(\frac{\nu_{i}-\mu_{j}}{2\pi}+\frac{m_{2}-m_{1}+m_{3}}{2}\right)\right)s_{b}\left(\frac{iQ}{4}-\left(\frac{\nu_{i}-\mu_{j}}{2\pi}+\frac{m_{2}-m_{1}-m_{3}}{2}\right)\right)\right\},\label{eq:zABJM-Gen}
\end{align}
where $Q=b+b^{-1}$. The mass parameter $m_{1}$ corresponds to $\mathrm{U}\left(1\right)$ and $m_{2}+m_{3}$, $m_{2}-m_{3}$ correspond to the Cartans of each factor in $\mathrm{SU}\left(2\right)\times\mathrm{SU}\left(2\right)$, respectively.
The matrix model is invariant under $b\leftrightarrow b^{-1}$
(note that the double sine function $s_b$ is invariant as \eqref{eq:DSF-props}),
which reflects the fact that the squashed three sphere is invariant under this exchange $S_b^3=S_{b^{-1}}^3$.
When $m_{3}=0$ (or $m_{2}=0$), the integrand can be written in terms of the $\mathcal{D}_{b}$ function defined in \eqref{eq:D-Def} as
\begin{align}
&Z_{k,b}^{\mathrm{ABJM}}\left(N;m_{1},m_{2},0\right)\nonumber \\
&=\frac{1}{\left(N!\right)^{2}}\int_{\mathbb{R}}\prod_{i}^{N}\frac{d\mu_{i}}{2\pi}\frac{d\nu_{i}}{2\pi}e^{-\frac{ik}{4\pi}\sum_{i}^{N}\left(\mu_{i}^{2}-\nu_{i}^{2}\right)}\prod_{i<j}^{N}\prod_{\pm}\left\{\left(2\sinh\frac{b^{\pm 1}\left(\mu_{i}-\mu_{j}\right)}{2}\right)\left(2\sinh\frac{b^{\pm 1}\left(\nu_{i}-\nu_{j}\right)}{2}\right)\right\}\nonumber \\
 & \quad\quad\times\prod_{i,j}^{N}\left\{\mathcal{D}_{b}\left(\frac{\mu_{i}-\nu_{j}}{2\pi}+\frac{m_{2}+m_{1}}{2}\right)\mathcal{D}_{b}\left(\frac{\nu_{i}-\mu_{j}}{2\pi}+\frac{m_{2}-m_{1}}{2}\right)\right\}.\label{eq:zABJM-m30}
\end{align}

The $\mathcal{N}=4$ $\mathrm{U}\left(N\right)$ SYM theory with an adjoint hypermultiplet and a fundamental hypermultiplet admits the FI deformation $\zeta$ and the mass deformation $m$ for the adjoint hypermultiplet.
The matrix model is
\begin{align}
Z_{b}\left(N;\zeta,m\right) & =\frac{1}{N!}\int_{\mathbb{R}}\prod_{i}^{N}\frac{d\mu_{i}}{2\pi}e^{i\zeta\sum_{i}^{N}\mu_{i}}\prod_{i<j}^{N}\left\{\left(2\sinh\frac{b\left(\mu_{i}-\mu_{j}\right)}{2}\right)\left(2\sinh\frac{\mu_{i}-\mu_{j}}{2b}\right)\right\}\nonumber \\
 & \quad\quad\times\prod_{i,j}^{N}\mathcal{D}_{b}\left(\frac{\mu_{i}-\mu_{j}}{2\pi}+m\right)\prod_{i}^{N}\mathcal{D}_{b}\left(\frac{\mu_{i}}{2\pi}\right).\label{eq:zADHM-Gen}
\end{align}
Here for the $\mathcal{N}=4$ SYM theory we omit the label ``SYM''.
The matrix model is again invariant under $b\leftrightarrow b^{-1}$.
Hence in the remaining part of this article we assume $b\geq1$.

The ABJM theory with $k=1$ is IR dual to the $\mathcal{N}=4$ SYM theory.
They share $b$ and $N$, while when $m_3=0$, the parameter correspondence between $\left(m_{1},m_{2}\right)$ and $\left(\zeta,m\right)$ is
\begin{equation}
\zeta=\frac{1}{2}\left(m_{2}+m_{1}\right),\quad
m=\frac{1}{2}\left(m_{2}-m_{1}\right).\label{eq:Duality-Para}
\end{equation}
See appendix \ref{sec:Duality} for the check of these relations when $b=1$ or $N=1$.\footnote{
The duality relation including the FI deformation and the mass deformations was also studied numerically for general values of $b\ge 1$ and $N\le 3$ in \cite{Thull:2022lif}.
}

Note that the partition function of the ABJM theory \eqref{eq:zABJM-m30} has the following symmetries
\begin{align}
& Z_{k,b}^{\mathrm{ABJM}}\left(N;m_{1},m_{2},0\right)=Z_{k,b}^{\mathrm{ABJM}}\left(N;-m_{2},-m_{1},0\right),\nonumber\\
& Z_{k,b}^{\mathrm{ABJM}}\left(N;m_{1},m_{2},0\right)=Z_{k,b}^{\mathrm{ABJM}}\left(N;m_{2},m_{1},0\right),\nonumber\\
& Z_{k,b}^{\mathrm{ABJM}}\left(N;m_{1},m_{2},0\right)=Z_{k,b}^{\mathrm{ABJM}}\left(N;-m_{1},m_{2},0\right).
\end{align}
Correspondingly, the partition function of the $\mathcal{N}=4$ SYM theory \eqref{eq:zADHM-Gen} has the following symmetries
\begin{align}
& Z_b(N;\zeta,m)=Z_b(N;-\zeta,m),\quad
Z_b(N;\zeta,m)=Z_b(N;\zeta,-m),\nonumber\\
& Z_b(N;\zeta,m)=Z_b(N;m,\zeta).
\end{align}
The first two symmetries are not manifest in the ABJM partition function \eqref{eq:zABJM-m30} but manifest in the partition function of the $\mathcal{N}=4$ SYM theory \eqref{eq:zADHM-Gen}.
On the other hand, the third symmetry is manifest in \eqref{eq:zABJM-m30} but not obvious from the expression \eqref{eq:zADHM-Gen}.
In the $\mathcal{N}=4$ SYM theory, the third symmetry follows from the fact that as an $\mathcal{N}=4$ supersymmetric field theory the $\mathcal{N}=4$ SYM theory is mirror to itself \cite{Kapustin:2010xq}.
In section \ref{sec:NumericComp} we show in several examples that the symmetry is indeed satisfied in the $\mathcal{N}=4$ SYM theory.

\subsection{Large $N$ conjecture\label{subsec:LargeN-Airy}}

In this section we review a proposal of the large $N$ behavior of the partition function.
It was conjectured that, even after deformed by $b$ and $m_{1,2,3}$, the form of the exact large $N$ perturbative expansion of the ABJM matrix model is still the Airy form as \cite{Bobev:2022jte,Hristov:2022lcw,Bobev:2022eus}
\begin{equation}
Z_{k,b}^{\mathrm{ABJM}\left(\mathrm{pert}\right)}\left(N;\bm{m}\right)=e^{A_{k,b}\left(\bm{m}\right)}C_{k,b}\left(\bm{m}\right)^{-\frac{1}{3}}\mathrm{Ai}\left[C_{k,b}\left(\bm{m}\right)^{-\frac{1}{3}}\left(N-B_{k,b}\left(\bm{m}\right)\right)\right],\label{eq:zABJMpert}
\end{equation}
where
\begin{align}
 & B_{k,b}\left(\bm{m}\right)=\left\{ -\frac{1}{12}\sum_{a=1}^{4}\Delta_{a}^{-1}+\frac{1-\frac{1}{4}\sum_{a=1}^{4}\Delta_{a}^{2}}{3\prod_{a=1}^{4}\Delta_{a}}\left(b+\frac{1}{b}\right)^{-2}\right\} \frac{1}{k}+\frac{k}{24},\nonumber \\
 & C_{k,b}\left(\bm{m}\right)=\frac{2}{\prod_{a=1}^{4}\Delta_{a}}\left(b+\frac{1}{b}\right)^{-4}\frac{1}{\pi^{2}k}.\label{eq:BC-ABJM}
\end{align}
Here $\Delta_{a}$ ($a=1,2,3,4$) are defined as
\begin{align}
\Delta_{1} & =\frac{1}{2}-i\frac{m_{1}+m_{2}+m_{3}}{b+b^{-1}},\quad\Delta_{2}=\frac{1}{2}-i\frac{m_{1}-m_{2}-m_{3}}{b+b^{-1}},\nonumber \\
\Delta_{3} & =\frac{1}{2}+i\frac{m_{1}+m_{2}-m_{3}}{b+b^{-1}},\quad\Delta_{4}=\frac{1}{2}+i\frac{m_{1}-m_{2}+m_{3}}{b+b^{-1}}.\label{eq:Delta-ABJM}
\end{align}
On the other hand, $A_{k,b}\left(\bm{m}\right)$ is known only for particular cases.
When $b=1$, $A_{k,1}\left(m_1,m_2,0\right)$ was obtained with the help of the Fermi gas formalism and matches the exact results for finite $N$ \cite{Nosaka:2015iiw}.
When $b=\sqrt{3}$, $A_{1,\sqrt{3}}\left(0,0,0\right)$ was obtained again with the help of the Fermi gas formalism \cite{Hatsuda:2016uqa}.
Although $A_{k,b}\left(0,0,0\right)$ for general parameters was also conjectured in \cite{Bobev:2022eus}, in the same paper the authors pointed out a discrepancy
with the result in \cite{Hatsuda:2016uqa} for $b=\sqrt{3}$.
In this sense, $A_{k,b}\left(\bm{m}\right)$ for general $b$ is not known.

The above result for the ABJM theory can be translated to that for the $\mathcal{N}=4$ SYM theory by using the duality.
Because the parameter correspondence of the duality is \eqref{eq:Duality-Para}, the perturbative part of the SYM partition function is expected to be
\begin{align}
Z_{b}^{\left(\mathrm{pert}\right)}\left(N;\zeta,m\right)
=Z_{k=1,b}^{\mathrm{ABJM}\left(\mathrm{pert}\right)}\left(N;\zeta-m,\zeta+m,0\right).\label{eq:ADHMpert-ABJM}
\end{align}
In terms of the parameters of the SYM, $B_{k,b}\left(\bm{m}\right)$ and $C_{k,b}\left(\bm{m}\right)$ in \eqref{eq:BC-ABJM} are written as
\begin{align}
 & B_{1,b}\left(\zeta-m,\zeta+m,0\right)=-\frac{1}{12}\sum_{a=1}^{4}\left(\Delta_{a}'\right)^{-1}+\frac{1-\frac{1}{4}\sum_{a=1}^{4}\left(\Delta_{a}'\right)^{2}}{3\prod_{a=1}^{4}\Delta_{a}'}\left(b+\frac{1}{b}\right)^{-2}+\frac{1}{24},\nonumber \\
 & C_{1,b}\left(\zeta-m,\zeta+m,0\right)=\frac{2}{\prod_{a=1}^{4}\Delta_{a}'}\left(b+\frac{1}{b}\right)^{-4}\frac{1}{\pi^{2}},\label{eq:BC-ADHM}
\end{align}
with
\begin{align}
\Delta_{1}'=\frac{1}{2}-i\frac{2\zeta}{b+b^{-1}},
\quad\Delta_{2}'=\frac{1}{2}+i\frac{2m}{b+b^{-1}},
\quad\Delta_{3}'=\frac{1}{2}+i\frac{2\zeta}{b+b^{-1}},
\quad\Delta_{4}'=\frac{1}{2}-i\frac{2m}{b+b^{-1}}.
\label{eq:Delta-ADHM}
\end{align}

\section{Exact large $N$ expansion\label{sec:LargeN}}

In the previous section we have reviewed the conjecture of the Airy form for the exact large $N$ expansion of the ABJM matrix model and the SYM matrix model.
In this section we give an analytic result for the exact large $N$ expansion of the SYM matrix model with arbitrary $b$.
In section \ref{subsec:FGF}, we apply the Fermi gas formalism for odd $b^2$.
Then, in section \ref{subsecLargeN}, we obtain the large $N$ result by using the Fermi gas partition function.

\subsection{Fermi gas formalism\label{subsec:FGF}}

In this section we show that the Fermi gas formalism is applicable for arbitrary odd $b^2$ if we adjust the mass parameter.
Before achieving this in section \ref{subsec:FGF-bGen}, in sections \ref{subsec:FGF-b1} and \ref{subsec:FGF-b3} we review the Fermi gas formalism for $b^{2}=1$ and $b^{2}=3$ cases.
A technical reason why we focus on odd $b^2$ case is that the integrand of the matrix model is drastically simplified thanks to the formula for the $\mathcal{D}_b$ function \eqref{eq:D-bOdd}.

We first state generalities of the Fermi gas formalism, which we use throughout this section.
The Fermi gas system is described with ordinary quantum mechanics. Let $\hat{x}$ and $\hat{p}$ be position and momentum operators satisfying $\left[\hat{x},\hat{p}\right]=i\hbar$.
We normalize the position eigenvector $\ket x$ as $\braket{x_{1}|x_{2}}=\delta\left(x_{1}-x_{2}\right)$.
The operators $\hat{x}$ and $\hat{p}$ satisfy the following similarity transformation formulas and the Baker–Campbell–Hausdorff formula
\begin{align}
&e^{-\frac{ia}{\hbar}\hat{x}}f\left(\hat{p}\right)e^{\frac{ia}{\hbar}\hat{x}}=f\left(\hat{p}+a\right),\quad e^{-\frac{ia}{\hbar}\hat{p}}f\left(\hat{x}\right)e^{\frac{ia}{\hbar}\hat{p}}=f\left(\hat{x}-a\right), \nonumber \\
&e^{\hat{x}}e^{\hat{p}}=e^{\frac{1}{2}i\hbar}e^{\hat{x}+\hat{p}}=e^{i\hbar}e^{\hat{p}}e^{\hat{x}}.\label{eq:OpSim1}
\end{align}
The Fermi gas formalism is a rewriting of the matrix model into the partition function of an ideal Fermi gas
\begin{equation}
Z_{b}\left(N;\zeta,m\right)=\frac{1}{N!}\int_{\mathbb{R}}\prod_{i}^{N}d\mu_{i}\det\left(\left[\braket{\mu_{i}|\hat{\rho}_{b}\left(\zeta,m\right)|\mu_{j}}\right]_{i,j}^{N\times N}\right).\label{eq:MM-FGS}
\end{equation}
Here the operator $\hat{\rho}_{b}$ is called the density matrix.
Note that the partition function of the Fermi gas is invariant under a similarity transformation of the density matrix, which plays an important role in section \ref{subsec:FGF-b3} and section \ref{subsec:FGF-bGen}.

It was discovered for the ABJM theory and the $\mathcal{N}=4$ circular quiver gauge theories (with various rank deformations) that the inverse of the density matrix
\begin{equation}
\hat{{\cal O}}_{b}\left(\zeta,m\right)=\hat{\rho}_{b}\left(\zeta,m\right)^{-1},\label{eq:DM-QC}
\end{equation}
becomes a finite sum of exponential functions of ${\hat x}$ and ${\hat p}$ \cite{Marino:2011eh,Marino:2015ixa,Kashaev:2015wia,Hatsuda:2016uqa,Kubo:2020qed}, and hence it is called a ``quantum curve''.
The spectral problem and the large $N$ behavior of the quantum curve have been studied in the context of the topological string (see, e.g.~\cite{Kallen:2013qla,Huang:2014eha,Kallen:2014lsa,Grassi:2014zfa,Grassi:2014uua,Marino:2015ixa,Hatsuda:2015oaa,Kashaev:2015wia,Codesido:2015dia,Moriyama:2020lyk}).
In order to utilize these results, our goal in the following sections is to find a Fermi gas formalism for the partition functions where the inverse of the density matrices are written in the ``curve'' form.

\subsubsection{$b^{2}=1$ case\label{subsec:FGF-b1}}

As a warm-up, in this section we review the Fermi gas formalism when $b=1$ \cite{Kapustin:2010xq,Marino:2011eh}.
In this case, the matrix model \eqref{eq:zADHM-Gen} is
\begin{equation}
Z_{1}\left(N;\zeta,m\right)=\frac{1}{N!}\int_{\mathbb{R}}\prod_{i}^{N}\frac{d\mu_{i}}{2\pi}e^{i\zeta\sum_{i}^{N}\mu_{i}}\frac{1}{\prod_{i}^{N}2\cosh\frac{\mu_{i}}{2}}\frac{\prod_{i<j}^{N}\left(2\sinh\frac{\mu_{i}-\mu_{j}}{2}\right)^{2}}{\prod_{i,j}^{N}2\cosh\frac{\mu_{i}-\mu_{j}+2\pi m}{2}}.\label{eq:MM-ADHM-Def}
\end{equation}
For applying the Fermi gas formalism, the first step is to use the Cauchy determinant formula
\begin{equation}
\frac{\prod_{i<j}^{N}2\sinh\frac{\alpha_{i}-\alpha_{j}}{2}\prod_{i<j}^{N}2\sinh\frac{\beta_{i}-\beta_{j}}{2}}{\prod_{i,j}^{N}2\cosh\frac{\alpha_{i}-\beta_{j}+c}{2}}=\det\left(\left[\frac{1}{2\cosh\frac{\alpha_{i}-\beta_{j}+c}{2}}\right]_{i,j}^{N\times N}\right).\label{eq:CauchyDetForm}
\end{equation}
Here $\left( \left[ a_{i,j} \right]_{i,j}^{N\times N} \right)$ denotes a $N \times N$ matrix whose $(i,j)$ element is $a_{i,j}$.
By using this formula, we can rewrite the last factor in the matrix model \eqref{eq:MM-ADHM-Def} as a determinant.
The second step is to rewrite the matrix elements in the operator formalism
\begin{equation}
\frac{1}{2\cosh\frac{\alpha-\beta+c}{2}}=\frac{1}{2\pi}\int_\mathbb{R} dp\frac{e^{\frac{i}{2\pi}p\left(\alpha-\beta+c\right)}}{2\cosh\frac{p}{2}}=2\pi\braket{\alpha|\frac{e^{\frac{ic}{2\pi}\hat{p}}}{2\cosh\frac{\hat{p}}{2}}|\beta}.\label{eq:Cosh-Op}
\end{equation}
Here we have chosen the normalization of the momentum operator ${\hat p}$ so that the canonical commutation relation is $\left[\hat{x},\hat{p}\right]=2\pi i$.
At the first equality, we have used a Fourier transformation formula for $\cosh^{-1}$.
After putting the remaining factors in the integrand into the determinant, the matrix model \eqref{eq:MM-ADHM-Def} finally reduces into the form of \eqref{eq:MM-FGS} with
\begin{equation}
\hat{\rho}_{1}\left(\zeta,m\right)=\frac{e^{i\zeta\hat{x}}}{2\cosh\frac{\hat{x}}{2}}\frac{e^{im\hat{p}}}{2\cosh\frac{\hat{p}}{2}}.\label{eq:DM-b1}
\end{equation}

The quantum curve \eqref{eq:DM-QC} in this case is then
\begin{equation}
\hat{{\cal O}}_{1}\left(\zeta,m\right)=e^{-im\hat{p}}\left(e^{\frac{\hat{p}}{2}}+e^{-\frac{\hat{p}}{2}}\right)e^{-i\zeta\hat{x}}\left(e^{\frac{\hat{x}}{2}}+e^{-\frac{\hat{x}}{2}}\right).\label{eq:QC-b1}
\end{equation}
Note that taking the inverse of the density matrix reverses the order of the operators, and hence the position operators are on the right side.

\subsubsection{$b^{2}=3$ case\label{subsec:FGF-b3}}

In this section we apply the Fermi gas formalism to the $b=\sqrt{3}$ case.
This was already done without deformations $\zeta=m=0$ in \cite{Marino:2015ixa,Hatsuda:2016uqa}, and the way of computation is almost the same also for $\zeta \neq 0$.

After rescaling the integration variables as $\mu_{i}\rightarrow\sqrt{3}\mu_{i}$, by using \eqref{eq:D-bOdd} the matrix model \eqref{eq:zADHM-Gen} becomes
\begin{align}
Z_{\sqrt{3}}\left(N;\zeta,m\right) & =\frac{1}{N!}\int_{\mathbb{R}}\prod_{i}^{N}\left(\frac{\sqrt{3}}{2\pi}d\mu_{i}\right)e^{i\sqrt{3}\zeta\sum_{i}^{N}\mu_{i}}\prod_{i}^{N}\mathcal{D}_{\sqrt{3}}\left(\frac{\sqrt{3}}{2\pi}\mu_{i}\right)\nonumber \\
 & \quad\quad\times\frac{
 \prod_{i<j}^{N}\left\{\left(2\sinh\frac{\mu_{i}-\mu_{j}}{2}\right)^{2}2\sinh\left(\frac{\mu_{i}-\mu_{j}}{2}+\frac{\pi i}{3}\right)2\sinh\left(\frac{\mu_{i}-\mu_{j}}{2}-\frac{\pi i}{3}\right)\right\}
 }{
 \prod_{i,j}^{N}\left\{2\cosh\left(\frac{\mu_{i}-\mu_{j}}{2}+\frac{\pi}{\sqrt{3}}m+\frac{\pi i}{6}\right)2\cosh\left(\frac{\mu_{i}-\mu_{j}}{2}+\frac{\pi}{\sqrt{3}}m-\frac{\pi i}{6}\right)\right\}
 }.\label{eq:Zb3cp1}
\end{align}
Here we have decomposed the factor $2\sinh\frac{3}{2}\left(\mu_{i}-\mu_{j}\right)$ in the numerator by using the formula
\begin{align}
2\sinh\left(Mx\right)
=i^{M-1}
\prod_{\ell=0}^{M-1}2\sinh\left(x-\frac{\ell}{M}\pi i\right).
\label{decomposesinhMx}
\end{align}
As is the case for $b=1$, the factor $(2\sinh\frac{\mu_i-\mu_j}{2})^2$ and the first factor in the denominator in the integrand \eqref{eq:Zb3cp1} can be combined into a Cauchy determinant \eqref{eq:CauchyDetForm}.
On the other hand, the Cauchy determinant formula is not applicable to the remaining factors due to the shift $\pm\frac{\pi i}{3}$ in the argument of the sinh factors in the numerator.
An important observation, however, is that these factors cancel with the remaining cosh factor in the denominator when we set $m=0$.
Note that this cancellation can occur because the SYM theory is a one-node quiver theory.
In other words, this type of the cancellation would not occur for multi-cut matrix models such as the ABJM matrix model since the arguments of the sine and cosine hyperbolic functions are different.
We then obtain
\begin{equation}
Z_{\sqrt{3}}\left(N;\zeta,0\right)=\frac{1}{N!}\int_{\mathbb{R}}\prod_{i}^{N}\frac{d\mu_{i}}{2\pi}e^{i\sqrt{3}\zeta\sum_{i}^{N}\mu_{i}}
\det\left(\left[\frac{1}{2\cosh\left(\frac{\mu_i-\mu_j}{2}+\frac{\pi i}{6}\right)}\right]_{i,j}^{N\times N}\right)
\prod_{i}^{N}\mathcal{D}_{\sqrt{3}}\left(\frac{\sqrt{3}}{2\pi}\mu_{i}\right).
\end{equation}
By using \eqref{eq:Cosh-Op}, we find the form \eqref{eq:MM-FGS}
\begin{equation}
Z_{\sqrt{3}}\left(N;\zeta,0\right)=\frac{1}{N!}\int_{\mathbb{R}}\prod_{i}^{N}d\mu_{i}\det\left(\left[\braket{\mu_{i}|e^{i\sqrt{3}\zeta\hat{x}}\frac{e^{-\frac{1}{6}\hat{p}}}{2\cosh\frac{\hat{p}}{2}}\mathcal{D}_{\sqrt{3}}\left(\frac{\sqrt{3}}{2\pi}\hat{x}\right)|\mu_{j}}\right]_{i,j}^{N\times N}\right).
\end{equation}
Here the commutation relation is
\begin{equation}
\left[\hat{x},\hat{p}\right]=2\pi i.
\end{equation}
Remember that the $\mathcal{D}_b$ function is written by a ratio of double sine functions by definition \eqref{eq:D-Def}.
Then, after an appropriate similarity transformation, we find a density matrix
\begin{equation}
\hat{\rho}_{\sqrt{3}}\left(\zeta,0\right)=e^{\frac{i\sqrt{3}\zeta}{2}\hat{x}}s_{\sqrt{3}}\left(\frac{\sqrt{3}}{2\pi}\hat{x}+\frac{i}{\sqrt{3}}\right)\frac{e^{-\frac{1}{6}\hat{p}}}{2\cosh\frac{\hat{p}}{2}}\frac{e^{\frac{i\sqrt{3}\zeta}{2}\hat{x}}}{s_{\sqrt{3}}\left(\frac{\sqrt{3}}{2\pi}\hat{x}-\frac{i}{\sqrt{3}}\right)}.
\end{equation}

The quantum curve \eqref{eq:DM-QC} in this case is then
\begin{equation}
\hat{{\cal O}}_{\sqrt{3}}\left(\zeta,0\right)=e^{-\frac{i\sqrt{3}\zeta}{2}\hat{x}}s_{\sqrt{3}}\left(\frac{\sqrt{3}}{2\pi}\hat{x}-\frac{i}{\sqrt{3}}\right)\left(e^{\frac{2}{3}\hat{p}}+e^{-\frac{1}{3}\hat{p}}\right)\frac{e^{-\frac{i\sqrt{3}\zeta}{2}\hat{x}}}{s_{\sqrt{3}}\left(\frac{\sqrt{3}}{2\pi}\hat{x}+\frac{i}{\sqrt{3}}\right)}.
\end{equation}
For obtaining a ``curve'' form, we move the double sine function on the denominator to the left side beyond the momentum operator by using \eqref{eq:OpSim1}.
The double sine function crosses $e^{\frac{2}{3}\hat{p}}$ or $e^{-\frac{1}{3}\hat{p}}$, and for the first term $e^{\frac{2}{3}\hat{p}}$ a cancellation occurs between the double sine functions on the numerator and the denominator, while for the second term $e^{-\frac{1}{3}\hat{p}}$ the double sine functions are simplified by \eqref{eq:DSF-Cosh}.
As a result we obtain
\begin{equation}
\hat{{\cal O}}_{\sqrt{3}}\left(\zeta,0\right)=e^{-i\sqrt{3}\zeta\hat{x}+\frac{2}{3}\hat{p}}+e^{\left(\frac{3}{2}-i\sqrt{3}\zeta\right)\hat{x}-\frac{1}{3}\hat{p}}+e^{-\left(\frac{3}{2}+i\sqrt{3}\zeta\right)\hat{x}-\frac{1}{3}\hat{p}}.\label{eq:QC-b3}
\end{equation}
In general, a quantum curve has redundancies in its expression.
One redundancy is to introduce new position and momentum operators $\hat{X},\hat{P}$, and by using this redundancy a quantum curve with three terms can be simplified, where two terms are $e^{\hat{X}}$ and $e^{\hat{P}}$.
For \eqref{eq:QC-b3}, we introduce new variables
\begin{equation}
\hat{X}=\left(\frac{3}{2}-i\sqrt{3}\zeta\right)\hat{x}-\frac{1}{3}\hat{p},\quad\hat{P}=-i\sqrt{3}\zeta\hat{x}+\frac{2}{3}\hat{p},
\end{equation}
so that the quantum curve becomes
\begin{equation}
\hat{{\cal O}}_{\sqrt{3}}\left(\zeta,0\right)=e^{\hat{X}}+e^{\hat{P}}+e^{-\mathfrak{m}\hat{X}-\mathfrak{n}\hat{P}},
\end{equation}
where
\begin{equation}
\mathfrak{m}=\frac{1+\sqrt{3}i\zeta}{1-\sqrt{3}i\zeta},\quad\mathfrak{n}=\frac{1}{1-\sqrt{3}i\zeta}.
\end{equation}
Here the commutation relation for the new variables is
\begin{equation}
\left[\hat{X},\hat{P}\right]=2\pi\left(1-i\sqrt{3}\zeta\right)i.
\end{equation}

\subsubsection{General odd integer $b^2$ \label{subsec:FGF-bGen}}

In this section we apply the Fermi gas formalism when $b^2=2n-1$ ($n\geq 2$).\footnote{
\label{footnoteb>1}
Note that our analysis is not valid for $b=\sqrt{2n-1}$ with $n=1$, where the matrix model \eqref{ADHMMMgeneraloddb2} substituted with $m_{b=1}=-i/2$ is divergent due to the factor in the denominator with $i=j$. 
For this reason, and since the large $N$ expansion for $b=1$ is already obtained without specifying the value of $m$, in this section we focus only on $b=\sqrt{2n-1}$ with $n\ge 2$.
}
After rescaling the integration variable as $\mu_{i}\rightarrow b\mu_{i}$, by using \eqref{eq:D-bOdd} the matrix model \eqref{eq:zADHM-Gen} becomes
\begin{align}
 Z_{b=\sqrt{2n-1}}\left(N;\zeta,m\right)
 & =\frac{1}{N!}\int_{\mathbb{R}}\prod_{i}^{N}\left(\frac{b}{2\pi}d\mu_{i}\right)e^{ib\zeta\sum_{i}^{N}\mu_{i}}\prod_{i}^{N}\mathcal{D}_{b}\left(\frac{b}{2\pi}\mu_{i}\right)\nonumber \\
 & \quad\quad\times\frac{
 \prod_{i<j}^{N}\left\{ \left(2\sinh\frac{\mu_{i}-\mu_{j}}{2}\right)^{2}
 \prod_{\ell=1}^{n-1}\prod_{\pm}
 2\sinh\left(\frac{\mu_{i}-\mu_{j}}{2}\pm\frac{\ell}{b^{2}}\pi i\right)\right\}
 }{ \prod_{i,j}^{N}\prod_{\ell=1}^{n}2\cosh\left(\frac{\mu_{i}-\mu_{j}}{2}+\frac{\pi}{b}m+\frac{\pi i}{b^{2}}\left(\frac{n+1}{2}-\ell\right)\right)
 }.
\label{ADHMMMgeneraloddb2}
\end{align}
Here we have decomposed the $2\sinh\frac{b^2}{2}\left(\mu_{i}-\mu_{j}\right)$ factor by using the formula \eqref{decomposesinhMx}.
As discussed in $b^{2}=3$ case, we need to cancel the sinh factors including $\pm \ell\pi i/b^{2}$.
An important observation is that the cancelation occurs when $m=m_b$, where
\begin{equation}
m_{b}=\frac{b^{2}-3}{4b}i.\label{eq:Mass-FGF}
\end{equation}
Namely, in this case
\begin{equation}
\left.\frac{
\prod_{i<j}^{N}\prod_{\ell=1}^{n-1}\prod_{\pm}
2\sinh\left(\frac{\mu_{i}-\mu_{j}}{2}\pm\frac{\ell}{b^{2}}\pi i\right)
}{
\prod_{i,j}^{N}\prod_{\ell=1}^{n-1}2\cosh\left(\frac{\mu_{i}-\mu_{j}}{2}+\frac{b^{2}-3}{4b^{2}}\pi i+\frac{\pi i}{b^{2}}\left(\frac{n+1}{2}-\ell\right)\right)
}\right|_{b^2=2n-1}
=\frac{1}{b^{N}}.
\end{equation}
With this adjustment the matrix model becomes
\begin{equation}
Z_{b}\left(N;\zeta,m_b\right)=\frac{1}{N!}\int_{\mathbb{R}}\prod_{i}^{N}\frac{d\mu_{i}}{2\pi}e^{ib\zeta\sum_{i}^{N}\mu_{i}}\frac{\prod_{i<j}^{N}\left(2\sinh\frac{\mu_{i}-\mu_{j}}{2}\right)^{2}}{\prod_{i,j}^{N}2\cosh\left(\frac{\mu_{i}-\mu_{j}}{2}+\frac{\pi i}{2b^{2}}\right)}\prod_{i}^{N}\mathcal{D}_{b}\left(\frac{b}{2\pi}\mu_{i}\right).
\end{equation}
Here we have used the fact that the hyperbolic cosine is parity even. By using \eqref{eq:CauchyDetForm} and \eqref{eq:Cosh-Op} in a similar way, we find the form \eqref{eq:MM-FGS}
\begin{equation}
Z_{b}\left(N;\zeta,m_b\right)=\frac{1}{N!}\int_{\mathbb{R}}\prod_{i}^{N}d\mu_{i}\det\left(\left[\braket{\mu_{i}|e^{ib\zeta\hat{x}}\frac{e^{-\frac{1}{2b^{2}}\hat{p}}}{2\cosh\frac{\hat{p}}{2}}\mathcal{D}_{b}\left(\frac{b}{2\pi}\hat{x}\right)|\mu_{j}}\right]_{i,j}^{N\times N}\right).
\end{equation}
Here the commutation relation is
\begin{equation}
\left[\hat{x},\hat{p}\right]=2\pi i.
\end{equation}
Remember that the $\mathcal{D}_b$ function is written by a ratio of double sine functions by definition \eqref{eq:D-Def}.
Then, after an appropriate similarity transformation, we find a density matrix
\begin{equation}
\hat{\rho}_{b}\left(\zeta,m_b\right)=e^{\frac{ib\zeta}{2}\hat{x}}s_{b}\left(\frac{b}{2\pi}\hat{x}+\frac{i}{4}Q\right)\frac{e^{-\frac{1}{2b^{2}}\hat{p}}}{2\cosh\frac{\hat{p}}{2}}\frac{e^{\frac{ib\zeta}{2}\hat{x}}}{s_{b}\left(\frac{b}{2\pi}\hat{x}-\frac{i}{4}Q\right)},
\end{equation}
where we remind that $Q=b+b^{-1}$.

The quantum curve \eqref{eq:DM-QC} in this case is then
\begin{equation}
\hat{{\cal O}}_{b}\left(\zeta,m_b\right)=e^{-\frac{ib\zeta}{2}\hat{x}}s_{b}\left(\frac{b}{2\pi}\hat{x}-\frac{i}{4}Q\right)\left(e^{\left(\frac{1}{2}+\frac{1}{2b^{2}}\right)\hat{p}}+e^{-\left(\frac{1}{2}-\frac{1}{2b^{2}}\right)\hat{p}}\right)\frac{e^{-\frac{ib\zeta}{2}\hat{x}}}{s_{b}\left(\frac{b}{2\pi}\hat{x}+\frac{i}{4}Q\right)}.
\end{equation}
For obtaining a ``curve'' form, we move the double sine function on denominator to left side beyond the momentum operator by using \eqref{eq:OpSim1} as we have done in the $b^2=3$ case.
The double sine function crosses $e^{\left(\frac{1}{2}+\frac{1}{2b^{2}}\right)\hat{p}}$ or $e^{-\left(\frac{1}{2}-\frac{1}{2b^{2}}\right)\hat{p}}$, and for the first term $e^{\left(\frac{1}{2}+\frac{1}{2b^{2}}\right)\hat{p}}$ a cancellation occurs between the double sine functions on numerator and denominator, while for the second term $e^{-\left(\frac{1}{2}-\frac{1}{2b^{2}}\right)\hat{p}}$ the double sine functions are simplified by \eqref{eq:DSF-Cosh}.
As a result we obtain
\begin{equation}
\hat{{\cal O}}_{b}\left(\zeta,m_b\right)=e^{-ib\zeta\hat{x}+\left(\frac{1}{2}+\frac{1}{2b^{2}}\right)\hat{p}}+e^{\left(\frac{b^{2}}{2}-ib\zeta\right)\hat{x}-\left(\frac{1}{2}-\frac{1}{2b^{2}}\right)\hat{p}}+e^{-\left(\frac{b^{2}}{2}+ib\zeta\right)\hat{x}-\left(\frac{1}{2}-\frac{1}{2b^{2}}\right)\hat{p}}.
\end{equation}
Following the $b^2=3$ case, we introduce new variables
\begin{equation}
\hat{X}=\left(\frac{b^{2}}{2}-ib\zeta\right)\hat{x}-\left(\frac{1}{2}-\frac{1}{2b^{2}}\right)\hat{p},\quad\hat{P}=-ib\zeta\hat{x}+\left(\frac{1}{2}+\frac{1}{2b^{2}}\right)\hat{p},
\end{equation}
so that the quantum curve is simplified to be
\begin{equation}
\hat{{\cal O}}_{b}\left(\zeta,m_b\right)=e^{\hat{X}}+e^{\hat{P}}+e^{-\mathfrak{m}\hat{X}-\mathfrak{n}\hat{P}},\label{eq:QC-gen_b}
\end{equation}
where
\begin{equation}
\mathfrak{m}=\frac{b+b^{-1}+4i\zeta}{b+b^{-1}-4i\zeta},\quad\mathfrak{n}=\frac{2\left(b-b^{-1}\right)}{b+b^{-1}-4i\zeta}.
\end{equation}
Here the commutation relation for the new variables is
\begin{equation}
\left[\hat{X},\hat{P}\right]=\frac{b^{2}+1-4ib\zeta}{2}\pi i.\label{eq:ComRel-gen_b}
\end{equation}

\subsection{Exact large $N$ expansion from Fermi gas system\label{subsecLargeN}}

In the previous section we applied the Fermi gas formalism to the matrix model and obtained the three-term quantum curve \eqref{eq:QC-gen_b}.
The three-term quantum curve appeared in the context of the topological string \cite{Aganagic:2011mi,Grassi:2014zfa}
as the quantized version of the mirror curve of the anti-canonical bundle of the weighted projective space $\mathbb{P}\left(1,\mathfrak{m},\mathfrak{n}\right)$.
Motivated by these results, the three-term quantum curve has been studied in \cite{Hatsuda:2015oaa}.
They found that the large $N$ expansion of the partition function of the three-term quantum curve is the Airy form \eqref{eq:zABJMpert}, and $A,B,C$ of the Airy function for $\mathbb{P}\left(1,\mathfrak{m},\mathfrak{n}\right)$ were obtained.
Namely, when the matrix model has the Fermi gas form \eqref{eq:MM-FGS} and the quantum curve defined by the inverse of the density matrix \eqref{eq:DM-QC} has the form
\begin{equation}
\hat{{\cal O}}_{\mathfrak{m},\mathfrak{n}}\left(\hat{x},\hat{p}\right)=e^{\hat{x}}+e^{\hat{p}}+e^{-\mathfrak{m}\hat{x}-\mathfrak{n}\hat{p}},\label{eq:P2QC}
\end{equation}
with the commutation relation
\begin{equation}
\left[\hat{x},\hat{p}\right]=i\hbar,\label{eq:P2ComRel}
\end{equation}
the exact large $N$ perturbative expansion of the matrix model was found to be the Airy form \eqref{eq:zABJMpert}, and $A,B,C$ were obtained
\begin{align}
A_{\mathfrak{m},\mathfrak{n}}\left(\hbar\right)&
=\frac{1}{4}\left[\cm\left(\frac{\hbar}{\pi}\right)+\cm\left(\frac{\mathfrak{m}\hbar}{\pi}\right)+\cm\left(\frac{\mathfrak{n}\hbar}{\pi}\right)-\cm\left(\frac{\left(\mathfrak{m}+\mathfrak{n}+1\right)\hbar}{\pi}\right)\right],\nonumber \\
B_{\mathfrak{m},\mathfrak{n}}\left(\hbar\right) & =\frac{\mathfrak{m}^{2}+\mathfrak{m}\mathfrak{n}+\mathfrak{n}^{2}+\mathfrak{m}+\mathfrak{n}+1}{12\mathfrak{m}\mathfrak{n}}\frac{\pi}{\hbar}-\frac{\mathfrak{m}+\mathfrak{n}+1}{48\pi}\hbar,\nonumber \\
C_{\mathfrak{m},\mathfrak{n}}\left(\hbar\right) & =\frac{\left(\mathfrak{m}+\mathfrak{n}+1\right)^{2}}{\mathfrak{m}\mathfrak{n}}\frac{1}{4\pi\hbar}.\label{eq:P2QC-ABC}
\end{align}
Here $\cm(z)$ is a constant map studied in \cite{Hanada:2012si,Hatsuda:2014vsa}
\begin{equation}
\cm\left(z\right)=\frac{2\zeta\left(3\right)}{\pi^{2}z}\left(1-\frac{z^{3}}{16}\right)+\frac{z^{2}}{\pi^{2}}\int_{0}^{\infty}dx\frac{x}{e^{zx}-1}\log\left(1-e^{-2x}\right).\label{eq:CM-Def}
\end{equation}

By comparing our result (\eqref{eq:QC-gen_b} and \eqref{eq:ComRel-gen_b}) with \eqref{eq:P2QC} and \eqref{eq:P2ComRel}, we find that in our case the parameters are
\begin{align}
\hbar =\frac{b^{2}+1-4ib\zeta}{2}\pi,
\quad \mathfrak{m} =\frac{b+b^{-1}+4i\zeta}{b+b^{-1}-4i\zeta},
\quad \mathfrak{n}=\frac{2\left(b-b^{-1}\right)}{b+b^{-1}-4i\zeta}.\label{para-ADHM-P2}
\end{align}
Notice that in \cite{Hatsuda:2015oaa}, \eqref{eq:P2QC-ABC} was obtained for $\hbar>0$, $\mathfrak{m}>0$ and $\mathfrak{n}>0$.
Although in our case the parameters are complex for non-zero $\zeta$, we expect that \eqref{eq:P2QC-ABC} holds for complex parameters.
By substituting \eqref{para-ADHM-P2} into \eqref{eq:P2QC-ABC}, we finally find that when $m=m_{b}$, the exact large $N$ expansion of the matrix model \eqref{eq:zADHM-Gen} up to the non-perturbative corrections is
\begin{equation}
Z_{b}^{\left(\mathrm{pert}\right)}\left(N;\zeta,m_b\right)=e^{A_{b}\left(\zeta,m_b\right)}C_{b}\left(\zeta,m_b\right)^{-\frac{1}{3}}{\mathrm{Ai}}\left[C_{b}\left(\zeta,m_b\right)^{-\frac{1}{3}}\left(N-B_{b}\left(\zeta,m_b\right)\right)\right],\label{eq:zADHMpert}
\end{equation}
with
\begin{align}
A_{b}\left(\zeta,m_{b}\right) & =\frac{1}{4}\left[\cm\left(\frac{b^{2}+1-4ib\zeta}{2}\right)+\cm\left(\frac{b^{2}+1+4ib\zeta}{2}\right)+\cm\left(b^{2}-1\right)-\cm\left(2b^{2}\right)\right],\nonumber \\
B_{b}\left(\zeta,m_{b}\right) & =-\frac{b^{8}+b^{6}\left(16\zeta^{2}+1\right)-b^{4}\left(16\zeta^{2}+23\right)+b^{2}\left(32\zeta^{2}+3\right)-6}{24\left(b^{2}-1\right)\left(b^{2}-4ib\zeta+1\right)\left(b^{2}+4ib\zeta+1\right)},\nonumber \\
C_{b}\left(\zeta,m_{b}\right) & =\frac{4b^{4}}{\pi^{2}\left(b^{2}-1\right)\left(b^{2}-4ib\zeta+1\right)\left(b^{2}+4ib\zeta+1\right)}.\label{eq:ABC-FGF}
\end{align}

Here we obtained the large $N$ partition function based on the Fermi gas analysis which we applied for odd integer $b^2$.
However, the final result \eqref{eq:ABC-FGF} has analytic expression of $b$.
Therefore, we expect that this expression also holds for general $b>1$.
In section \ref{subsec:NumCheck} we provide evidence of this prediction.

Now we compare our result with the conjecture \eqref{eq:BC-ADHM}.
In our case, the mass parameter is fixed to be $m=m_{b}$ defined in \eqref{eq:Mass-FGF}, and thus \eqref{eq:Delta-ADHM} becomes
\begin{align}
\Delta_{1} =\frac{1}{2}-i\frac{2\zeta}{b+b^{-1}},
\quad\Delta_{2}=\frac{2b^{-1}}{b+b^{-1}},
\quad\Delta_{3} =\frac{1}{2}+i\frac{2\zeta}{b+b^{-1}},
\quad\Delta_{4}=\frac{b-b^{-1}}{b+b^{-1}}.
\end{align}
By substituting this to the conjecture \eqref{eq:BC-ADHM}, we obtain $B_{b}\left(\zeta,m_{b}\right)$ and $C_{b}\left(\zeta,m_{b}\right)$ in \eqref{eq:ABC-FGF}.
Therefore, we could confirm for $k=1,m_3=0,(m_2-m_1)/2=m_{b}$ that the conjecture \eqref{eq:BC-ABJM} is accurate.
On the other hand, $A_b(\zeta,m_b)$ in \eqref{eq:ABC-FGF} is a new result.
As commented in section \ref{subsec:LargeN-Airy}, the conjecture for $A_{k,b}\left(0,0,0\right)$ in \cite{Bobev:2022eus} is inconsistent with the $b^2=3$ rigorous result in \cite{Hatsuda:2016uqa}, while our result involves the $b^2=3$ result, and thus it is consistent with the conjecture in \cite{Hatsuda:2016uqa}.
Because our derivation is also rigorous, and we will check it numerically in section \ref{subsec:NumCheck}, we expect that our result is correct.

Finally, we comment on the $b\rightarrow 1$ limit.
In this limit the parameter $A_{b}\left(\zeta,m_{b}\right)$ diverges.
(Note that $\cm(z)$ diverges in the limit $z\rightarrow 0$.)
This is consistent with the fact that the matrix model itself diverges in this limit as commented in footnote
\ref{footnoteb>1}.
Note that the same divergence occurs when we set $b=1$ first, where $A_1(\zeta,m)$ is obtained in \cite{Nosaka:2015iiw}, and then take the limit $m\rightarrow -\frac{i}{2}$.

\section{Exact partition function at finite $N$ \label{sec:NumericComp}}

In this section we study the partition function \eqref{eq:zADHM-Gen} for finite $N$.
For the squashing parameters $b=\sqrt{2n-1}$ with $n\in \mathbb{N}$, the partition function \eqref{eq:zADHM-Gen} simplifies due to the formula for $\mathcal{D}_{b=\sqrt{2n-1}}$ function \eqref{eq:D-bOdd} as
\begin{align}
&Z_{b=\sqrt{2n-1}}(N;\zeta,m)\nonumber \\
&=\frac{
(2n-1)^{\frac{N}{2}}
}{N!}
\int_\mathbb{R}\prod_{i}^N\frac{d\mu_i}{2\pi}
e^{i\sqrt{2n-1}\zeta\sum_{i}^N\mu_i}
\prod_{i<j}^N
2\sinh\frac{(2n-1)(\mu_i-\mu_j)}{2}
2\sinh\frac{\mu_i-\mu_j}{2}\nonumber \\
&\quad\times \prod_{i,j}^N\prod_{\ell=1}^n\frac{1}{2\cosh[\frac{\mu_i-\mu_j}{2}+\frac{\pi m}{\sqrt{2n-1}}+\frac{\pi i}{2n-1}(\frac{n+1}{2}-\ell)]}
\prod_{i}^N\prod_{\ell=1}^n\frac{1}{2\cosh[\frac{\mu_i}{2}+\frac{\pi i}{2n-1}(\frac{n+1}{2}-\ell)]}.
\label{Zbsimplified}
\end{align}
We see that the closed form expression for the partition function can be obtained by evaluating the integration \eqref{Zbsimplified} by the residue sum.
To proceed, let us write the partition function \eqref{Zbsimplified} schematically as
\begin{align}
Z_{b=\sqrt{2n-1}}(N;\zeta,m)=\frac{1}{N!}\int_\mathbb{R} \prod_{i}^N\frac{d\mu_i}{2\pi} 
{\cal I}\prod_{i<j}^N\prod_\pm {\cal J}_{ij\pm}\prod_i^N{\cal K}_i,
\end{align}
with
\begin{align}
&{\cal I}=
\Bigl(\frac{\sqrt{2n-1}}{\prod_{\ell=1}^n2\cosh[\frac{\pi m}{\sqrt{2n-1}}+\frac{\pi i}{2n-1}(\frac{n+1}{2}-\ell)]}\Bigr)^N
e^{i\sqrt{2n-1}\zeta\sum_{i}^N\mu_i}\nonumber \\
&\quad\times\prod_{i<j}^N2\sinh\frac{(2n-1)(\mu_i-\mu_j)}{2}
2\sinh\frac{\mu_i-\mu_j}{2},\nonumber \\
&{\cal J}_{ij\pm}=\prod_{\ell=1}^n\frac{1}{2\cosh[\frac{\mu_i-\mu_j}{2}\pm\frac{\pi m}{\sqrt{2n-1}}+\frac{\pi i}{2n-1}(\frac{n+1}{2}-\ell)]},\quad
{\cal K}_i=\prod_{\ell=1}^n\frac{1}{2\cosh[\frac{\mu_i}{2}+\frac{\pi i}{2n-1}(\frac{n+1}{2}-\ell)]}.
\end{align}
Note that the integrand has infinitely many poles labelled by integers, with respect to each integration variable $\mu_i$.
As we see below, however, at each step of the iterative integration the integrand enjoys a simple quasi-periodicity.
As a result the contribution from infinite tower of poles can be taken care of just by multiplying an appropriate overall factor to the contribution from the fundamental set of the poles.

Let us first consider the case with $N=1$
\begin{align}
Z_{b=\sqrt{2n-1}}(1;\zeta,m)=\int_\mathbb{R} \frac{d\mu_1}{2\pi}{\cal I}{\cal K}_1.
\label{ZbsimplifiedN1}
\end{align}
In this case, the integrand has the poles
\begin{align}
\mu_1=-\frac{2\pi i}{2n-1}\Bigl(\frac{n+1}{2}-\ell_1\Bigr)+\pi i+2\pi ia_1,\quad \ell_1=1,2,\cdots,n,\quad a_1\in\mathbb{Z},
\label{polesN1}
\end{align}
due to the factor ${\cal K}_1$.
However, since the integrand is quasi-periodic under $\mu_1\rightarrow \mu_1+2\pi i$
\begin{align}
{\cal I}{\cal K}_1\Bigr|_{\mu_1 \rightarrow \mu_1+2\pi i}= (-1)^ne^{-2\pi b\zeta}{\cal I}{\cal K}_1,
\end{align}
we can rewrite the integration \eqref{ZbsimplifiedN1} as
\begin{align}
Z_{b=\sqrt{2n-1}}(1;\zeta,m)=\frac{1}{1-(-1)^ne^{-2\pi \sqrt{2n-1}\zeta}}\int_\gamma \frac{d\mu_1}{2\pi}{\cal I}{\cal K}_1,
\end{align}
with $\gamma$ the counter-clockwise contour through
\begin{align}
(-\infty,\infty)
\cup (\infty,\infty+2\pi i)
\cup (-\infty+2\pi i,\infty+2\pi i)
\cup (-\infty,-\infty+2\pi i).
\end{align}
Now the integration \eqref{ZbsimplifiedN1} is evaluated only by summing over the poles \eqref{polesN1} with $a_1=0$ which are inside the contour $\gamma$
\begin{align}
Z_{b=\sqrt{2n-1}}(1;\zeta,m)=\frac{1}{1-(-1)^ne^{-2\pi \sqrt{2n-1}\zeta}}\sum_{\ell_1=1}^n
\Bigl[
{\cal I}\Bigr]_{\mu_1=-\frac{2\pi i}{2n-1}(\frac{n+1}{2}-\ell_1)+\pi i}
\prod_{\substack{\ell'=1\\ (\ell'\neq \ell_1)}}^n\frac{1}{2\sin\frac{\pi(\ell'-\ell_1)}{2n-1}},
\end{align}
where the last factor comes from the residue of ${\cal K}_1$.
As a result, we finally obtain
\begin{align}
Z_{b=\sqrt{2n-1}}(1;\zeta,m)=\prod_{\ell_1=1}^n\frac{1}{2\cosh[\frac{\pi \zeta}{\sqrt{2n-1}}+\frac{\pi i}{2n-1}(\frac{n+1}{2}-\ell_1)]2\cosh[\frac{\pi m}{\sqrt{2n-1}}+\frac{\pi i}{2n-1}(\frac{n+1}{2}-\ell_1)]}.
\label{ZN1b2oddbyresiduesum}
\end{align}
Note that for $N=1$ we can also evaluate the partition function \eqref{eq:zADHM-Gen} for general $b\in\mathbb{R}$ simply by using the Fourier transformation formula \eqref{eq:D-Fourier} as
\begin{align}
Z_b(1;\zeta,m)=\mathcal{D}_b(\zeta)\mathcal{D}_b(m).
\label{eq:zADHM-N1}
\end{align}
The result obtained by the residue sum \eqref{ZN1b2oddbyresiduesum} precisely coincides with this formula.

Next let us consider the case with $N=2$.
In this case one may consider the possible choices of the poles as
\begin{align}
&\text{choice 1: }\mu_1\text{: pole of }{\cal K}_1,\quad \mu_2\text{: pole of }{\cal K}_2,\nonumber \\
&\text{choice 2}_\pm\text{: }\mu_1\text{: pole of }{\cal K}_1,\quad \mu_2\text{: pole of }{\cal J}_{12\pm },
\label{N2choice1and2pm}
\end{align}
up to the permutations of the indices of $\mu_i$ which give identical contribution to the partition function.
Note that this combinatorial factor is precisely cancelled with the overall factor $\frac{1}{N!}$ of the partition function \eqref{eq:zADHM-Gen}.
Actually, the choice 1 does not contribute to the partition function since the residue, which is proportional to the factor ${\cal I}$, vanishes.
Hence it is sufficient to take into account only the choice 2, where the poles are given as
\begin{align}
&\mu_1=-\frac{2\pi i}{2n-1}\Bigl(\frac{n+1}{2}-\ell_1\Bigr)+\pi i+2\pi ia_1,\quad \ell_1=1,2,\cdots,n,\quad a_1\in\mathbb{Z},\nonumber \\
&\mu_2=\mu_1\pm \frac{2\pi m}{\sqrt{2n-1}}+\frac{2\pi i}{2n-1}\Bigl(\frac{n+1}{2}-\ell_2\Bigr)+\pi i+2\pi ia_2,\quad \ell_2=1,2,\cdots,n,\quad a_2\in\mathbb{Z}.
\label{polesN2}
\end{align}
Let us perform the integration over $\mu_2$ first and then perform the integration over $\mu_1$.
In the first step, the integrand ${\cal I}{\cal J}_{12+}{\cal J}_{12-}{\cal K}_1{\cal K}_2$ is quasi-periodic under the shift $\mu_2\rightarrow \mu_2+2\pi i$ as
\begin{align}
{\cal I}{\cal J}_{12+}{\cal J}_{12-}{\cal K}_1{\cal K}_2\Bigr|_{\mu_2\rightarrow \mu_2+2\pi i}
=(-1)^ne^{-2\pi \sqrt{2n-1}\zeta}{\cal I}{\cal J}_{12+}{\cal J}_{12-}{\cal K}_1{\cal K}_2,
\end{align}
while in the second step, the integrand left after performing the $\mu_2$-integration has a different quasi-periodicity
\begin{align}
(\text{integrand})|_{\mu_1\rightarrow \mu_2+2\pi i}
=e^{-4\pi \sqrt{2n-1}\zeta}(\text{integrand}).
\end{align}
Following the same idea as in the case of $N=1$, we can restrict the infinite towers of the poles \eqref{polesN2} to the ones with $a_1=a_2=0$ by introducing extra overall factors
$(1-e^{-4\pi \sqrt{2n-1}\zeta})^{-1}\times(1-(-1)^ne^{-2\pi \sqrt{2n-1}\zeta})^{-1}$.
As a result, the partition function $Z_{b=\sqrt{2n-1}}(2;\zeta,m)$ is given as
\begin{align}
Z_{b=\sqrt{2n-1}}(2;\zeta,m)
&=
\frac{1}{1-e^{-4\pi \sqrt{2n-1}\zeta}}
\frac{1}{1-(-1)^ne^{-2\pi \sqrt{2n-1}\zeta}}
\sum_{\ell_1,\ell_2=1}^n
\sum_\pm
\Bigl[
{\cal I}{\cal J}_{12,\mp}{\cal K}_2\Bigr]_{\text{pole}_\pm(\ell_1,\ell_2)}\nonumber \\
&\quad\times \prod_{\substack{\ell'=1\\ (\ell'\neq \ell_1)}}^n\frac{1}{2\sin\frac{\pi(\ell'-\ell_1)}{2n-1}}
\prod_{\substack{\ell'=1\\ (\ell'\neq \ell_2)}}^n\frac{1}{2\sin\frac{\pi(\ell_2-\ell')}{2n-1}},
\label{Zb2oddN2residuesum}
\end{align}
where $|_{\text{pole}_\pm(\ell_1,\ell_2)}$ stands for the substitution of \eqref{polesN2} with $a_1=a_2=0$ to $\mu_1$ and $\mu_2$.

For $N=3$ there are following three choices of the poles with non-vanishing residues:
\begin{align}
&\text{choice 1}_\pm\text{: }
\mu_1\text{: pole of }{\cal K}_1,\quad
\mu_2\text{: pole of }{\cal J}_{12\pm},\quad
\mu_3\text{: pole of }{\cal J}_{23\pm},
\label{N3choice1pm}
\end{align}
where the poles are given by
\begin{align}
&\mu_1=-\frac{2\pi i}{2n-1}\Bigl(\frac{n+1}{2}-\ell_1\Bigr)+\pi i+2\pi ia_1,\quad \ell_1=1,2,\cdots,n,\quad a_1\in\mathbb{Z},\nonumber \\
&\mu_2=\mu_1\pm \frac{2\pi m}{\sqrt{2n-1}}+\frac{2\pi i}{2n-1}\Bigl(\frac{n+1}{2}-\ell_2\Bigr)+\pi i+2\pi ia_2,\quad \ell_2=1,2,\cdots,n,\quad a_2\in\mathbb{Z},\nonumber \\
&\mu_3=\mu_2\pm \frac{2\pi m}{\sqrt{2n-1}}+\frac{2\pi i}{2n-1}\Bigl(\frac{n+1}{2}-\ell_3\Bigr)+\pi i+2\pi ia_3,\quad \ell_3=1,2,\cdots,n,\quad a_3\in\mathbb{Z},
\label{polesN31}
\end{align}
and
\begin{align}
&\text{choice 2: }
\mu_1\text{: pole of }{\cal K}_1,\quad
\mu_2\text{: pole of }{\cal J}_{12+},\quad
\mu_3\text{: pole of }{\cal J}_{13-},
\label{N3choice2}
\end{align}
where the poles are given by
\begin{align}
&\mu_1=-\frac{2\pi i}{2n-1}\Bigl(\frac{n+1}{2}-\ell_1\Bigr)+\pi i+2\pi ia_1,\quad \ell_1=1,2,\cdots,n,\quad a_1\in\mathbb{Z},\nonumber \\
&\mu_2=\mu_1+\frac{2\pi m}{\sqrt{2n-1}}+\frac{2\pi i}{2n-1}\Bigl(\frac{n+1}{2}-\ell_2\Bigr)+\pi i+2\pi ia_2,\quad \ell_2=1,2,\cdots,n,\quad a_2\in\mathbb{Z},\nonumber \\
&\mu_3=\mu_1-\frac{2\pi m}{\sqrt{2n-1}}+\frac{2\pi i}{2n-1}\Bigl(\frac{n+1}{2}-\ell_3\Bigr)+\pi i+2\pi ia_3,\quad \ell_3=1,2,\cdots,n,\quad a_3\in\mathbb{Z}.
\label{polesN32}
\end{align}
Due to the same argument as above, we can restrict the summation over the poles to those with $a_1=a_2=a_3=0$ by introducing the extra factor
$(1-(-1)^ne^{-6\pi \sqrt{2n-1}\zeta})^{-1}\times(1-e^{-4\pi \sqrt{2n-1}\zeta})^{-1}\times(1-(-1)^ne^{-2\pi \sqrt{2n-1}\zeta})^{-1}$
for the choice $1_\pm$ and the extra factor
$(1-(-1)^ne^{-6\pi \sqrt{2n-1}\zeta})^{-1}\times(1-(-1)^ne^{-2\pi \sqrt{2n-1}\zeta})^{-2}$
for the choice $2$.
As a result, we finally obtain
\begin{align}
Z_{b=\sqrt{2n-1}}(3;\zeta,m)&=
\frac{1}{1-(-1)^ne^{-6\pi \sqrt{2n-1}\zeta}}
\frac{1}{1-e^{-4\pi \sqrt{2n-1}\zeta}}
\frac{1}{1-(-1)^ne^{-2\pi \sqrt{2n-1}\zeta}}\nonumber \\
&\quad\times \sum_{\ell_1,\ell_2,\ell_3=1}^n
\sum_\pm
\Bigl[
{\cal I}{\cal J}_{12,\mp}\prod_{\pm'} {\cal J}_{13,\pm'}{\cal J}_{23,\mp}
{\cal K}_2 {\cal K}_3 \Bigr]_{\text{pole}1_\pm(\ell_1,\ell_2,\ell_3)}\nonumber \\
&\quad\times \prod_{\substack{\ell'=1\\ (\ell'\neq \ell_1)}}^n\frac{1}{2\sin\frac{\pi(\ell'-\ell_1)}{2n-1}}
\prod_{\substack{\ell'=1\\ (\ell'\neq \ell_2)}}^n\frac{1}{2\sin\frac{\pi(\ell_2-\ell')}{2n-1}}
\prod_{\substack{\ell'=1\\ (\ell'\neq \ell_3)}}^n\frac{1}{2\sin\frac{\pi(\ell_3-\ell')}{2n-1}}\nonumber \\
&+
\frac{1}{1-(-1)^ne^{-6\pi \sqrt{2n-1}\zeta}}
\Bigl(
\frac{1}{1-(-1)^ne^{-2\pi \sqrt{2n-1}\zeta}}
\Bigr)^2\nonumber \\
&\quad\times \sum_{\ell_1,\ell_2,\ell_3=1}^n
\Bigl[
e^{ib\zeta\sum_i\mu_i}{\cal I}_1{\cal J}_{12,-}{\cal J}_{13,+}\prod_\pm {\cal J}_{23,\pm}
{\cal K}_2 {\cal K}_3 \Bigr]_{\text{pole}2(\ell_1,\ell_2,\ell_3)}\nonumber \\
&\quad\times \prod_{\substack{\ell'=1\\ (\ell'\neq \ell_1)}}^n\frac{1}{2\sin\frac{\pi(\ell'-\ell_1)}{2n-1}}
\prod_{\substack{\ell'=1\\ (\ell'\neq \ell_2)}}^n\frac{1}{2\sin\frac{\pi(\ell_2-\ell')}{2n-1}}
\prod_{\substack{\ell'=1\\ (\ell'\neq \ell_3)}}^n\frac{1}{2\sin\frac{\pi(\ell_3-\ell')}{2n-1}},
\label{Zb2oddN3residuesum}
\end{align}
where $\text{pole}1_\pm(\ell_1,\ell_2,\ell_3)$ and $\text{pole}2(\ell_1,\ell_2,\ell_3)$ stand respectively for the substitution of \eqref{polesN31} and \eqref{polesN32} with $a_1=a_2=a_3=0$.

At least for $N\le 3$, one may organize the contributions from different poles as those associated with the Young diagrams with $|\lambda|=N$, as was done for $b=1$ in \cite{Gaiotto:2019mmf}, as
\begin{align}
&N=1\rightarrow \lambda=\ydiagram{1},\nonumber \\
&N=2,\text{ pole: choice 2}_+\text{ in \eqref{N2choice1and2pm}}\rightarrow \lambda=\ydiagram{2},\nonumber \\
&N=2,\text{ pole: choice 2}_-\text{ in \eqref{N2choice1and2pm}}\rightarrow \lambda=\ydiagram{1,1},\nonumber \\
&N=3,\text{ pole: choice 1}_+\text{ in \eqref{N3choice1pm}}\rightarrow \lambda=\ydiagram{3},\nonumber \\
&N=3,\text{ pole: choice 2}\text{ in \eqref{N3choice2}}\rightarrow \lambda=\ydiagram{2,1},\nonumber \\
&N=3,\text{ pole: choice 1}_-\text{ in \eqref{N3choice1pm}}\rightarrow \lambda=\ydiagram{1,1,1}.
\end{align}
As a result, we find that the partition function is written as
\begin{align}
&Z_{b=\sqrt{2n-1}}(1;\zeta,m)=z_{b=\sqrt{2n-1},\ydiagram{1}}(\zeta,m),\\
&Z_{b=\sqrt{2n-1}}(2;\zeta,m)=
Rz_{b=\sqrt{2n-1},\ydiagram{2}}(\zeta,m)
+R^{-1}z_{b=\sqrt{2n-1},\ydiagram{1,1}}(\zeta,m),\label{ZN2residuefinal} \\
&Z_{b=\sqrt{2n-1}}(3;\zeta,m)=
R^3z_{b=\sqrt{2n-1},\ydiagram{3}}(\zeta,m)
+z_{b=\sqrt{2n-1},\ydiagram{2,1}}(\zeta,m)
+R^{-3}z_{b=\sqrt{2n-1},\ydiagram{1,1,1}}(\zeta,m),
\end{align}
where $R=ie^{2\pi i\zeta m}$ and $z_{b=\sqrt{2n-1},\lambda}(\zeta,m)$ are some rational functions of $e^{\frac{\pi \zeta}{\sqrt{2n-1}}}$ and $e^{\frac{\pi m}{\sqrt{2n-1}}}$ ($z_{b,\ydiagram{1}}(\zeta,m)$ is simply given by $\mathcal{D}_b(\zeta)\mathcal{D}_b(m)$).
We also find that $z_{b=\sqrt{2n-1},\lambda}(\zeta,m)$ satisfy
\begin{align}
&z_{b=\sqrt{2n-1},\ydiagram{1,1}}(\zeta,m)=-z_{b=\sqrt{2n-1},\ydiagram{2}}(\zeta,-m),\quad
z_{b=\sqrt{2n-1},\ydiagram{1,1,1}}(\zeta,m)=-z_{b=\sqrt{2n-1},\ydiagram{3}}(\zeta,-m),\nonumber \\
&z_{b=\sqrt{2n-1},\ydiagram{2,1}}(\zeta,-m)=z_{b=\sqrt{2n-1},\ydiagram{2,1}}(\zeta,m),\nonumber \\
&z_{b=\sqrt{2n-1},\lambda}(\zeta,m)=z_{b=\sqrt{2n-1},\lambda}(m,\zeta).
\end{align}
The last property manifests the fact that the $\mathcal{N}=4$ $\mathrm{U}\left(N\right)$ SYM theory with an adjoint hypermultiplet and a fundamental hypermultiplet is mirror to itself.

In particular, for $b=\sqrt{3}$ we have\footnote{
We were informed that the exact partition function for $N=2$ and $b=\sqrt{3}$ obtained from \eqref{ZN2residuefinal} with \eqref{zydiagram} agrees with the result of a direct numerical evaluation of the integration \eqref{eq:zADHM-Gen} performed in \cite{Thull:2022lif}.
}
\begin{align}
&z_{b=\sqrt{3},\ydiagram{2}}(\zeta,m)\nonumber \\
&=\frac{
\sinh\frac{\pi \zeta}{\sqrt{3}}
\sinh\frac{\pi m}{\sqrt{3}}
}{
4
\sinh(\sqrt{3}\pi \zeta)
\sinh(\sqrt{3}\pi m)
\sinh(2\sqrt{3} \pi \zeta)
\sinh(2\sqrt{3} \pi m)
}
\biggl[\sqrt{3} \Bigl(-1 - 2 \cosh\frac{2 \pi m}{\sqrt{3}} \cosh\frac{2 \pi \zeta}{\sqrt{3}}\Bigr)\nonumber \\
&\quad\quad + 2 i \sinh\frac{2 \pi m}{\sqrt{3}} \sinh\frac{2 \pi \zeta}{\sqrt{3}}\biggr],\nonumber \\
&z_{b=\sqrt{3},\ydiagram{3}}(\zeta,m)\nonumber \\
&=
\frac{
\sinh\frac{\pi m}{\sqrt{3}}
\sinh\frac{\pi \zeta}{\sqrt{3}}
}{
16
\sinh(\sqrt{3}    \pi m)
\sinh(2 \sqrt{3}  \pi m)
\sinh( 3 \sqrt{3} \pi m)
\sinh(\sqrt{3} \pi \zeta)
\sinh( 2 \sqrt{3} \pi \zeta)
\sinh(3 \sqrt{3} \pi \zeta)
}\nonumber \\
&\quad\times \biggl[
2 i \Bigl(
2 \cosh\frac{\pi m}{\sqrt{3}} \cosh\frac{\pi \zeta}{\sqrt{3}}
+ 3 \cosh(\sqrt{3} \pi m) \cosh(\sqrt{3} \pi \zeta)
+ 2 \cosh\frac{5 \pi m}{\sqrt{3}} \cosh\frac{5 \pi \zeta}{\sqrt{3}}\nonumber \\
&\quad\quad + \cosh\frac{5 \pi m}{\sqrt{3}} \cosh\frac{\pi \zeta}{\sqrt{3}}
+ \cosh\frac{\pi m}{\sqrt{3}} \cosh\frac{5 \pi \zeta}{\sqrt{3}}
\Bigr)
- 2 \sqrt{3}
\sinh\frac{\pi m}{\sqrt{3}} \sinh\frac{\pi \zeta}{\sqrt{3}}
\Bigl(
1\nonumber \\
&\quad\quad + 4 \cosh\frac{2 \pi m}{\sqrt{3}} \cosh\frac{2 \pi \zeta}{\sqrt{3}}
+ 2 \cosh\frac{4 \pi m}{\sqrt{3}}
+ 2 \cosh\frac{4 \pi \zeta}{\sqrt{3}}
\Bigr)
\biggr],\nonumber \\
&z_{b=\sqrt{3},\ydiagram{2,1}}(\zeta,m)\nonumber \\
&=
\frac{
\sinh\frac{ \pi m}{\sqrt{3}}
\sinh\frac{\pi \zeta}{\sqrt{3}}
}{
16
\sinh( \sqrt{3} \pi m)^2
\sinh(3 \sqrt{3} \pi m)
\sinh( \sqrt{3} \pi \zeta)^2
\sinh(3 \sqrt{3} \pi \zeta)
}
\Big(3
+ 4 \cosh\frac{2 \pi m}{\sqrt{3}} \cosh\frac{2 \pi \zeta}{\sqrt{3}}\nonumber \\
&\quad\quad + 4 \cosh\frac{4 \pi m}{\sqrt{3}} \cosh\frac{4 \pi \zeta}{\sqrt{3}}
- 6 \cosh\frac{2 \pi m}{\sqrt{3}}
- 6 \cosh\frac{2 \pi \zeta}{\sqrt{3}}
- 4 \cosh\frac{4 \pi m}{\sqrt{3}} \cosh\frac{2 \pi \zeta}{\sqrt{3}}\nonumber \\
&\quad\quad - 4 \cosh\frac{2 \pi m}{\sqrt{3}} \cosh\frac{4 \pi \zeta}{\sqrt{3}}
\Bigr).
\label{zydiagram}
\end{align}

\section{Numerical checks\label{subsec:NumCheck}}

In this section we perform two numerical tests and a numerical analysis of the results obtained in sections \ref{sec:LargeN} and \ref{sec:NumericComp}.
In section \ref{sec:LargeN} we have applied the Fermi gas formalism when $b^2$ is odd and $m$ is tuned to $m=m_b=\frac{b^2-3}{4b}i$ \eqref{eq:Mass-FGF}, and by using the quantum curve we have obtained the all order $1/N$ corrections as the Airy function \eqref{eq:zADHMpert} with \eqref{eq:ABC-FGF}.
Although the Airy function is rigorously obtained only when $b^2$ is an odd integer, the coefficients \eqref{eq:ABC-FGF} have analytic expressions with respect to $b$.
Hence, as commented in section \ref{subsecLargeN}, it would be natural to expect that the Airy function also holds for arbitrary $b>1$.
One test we perform is this, and we show that the numerical result strongly suggests that this is true.

\begin{figure}
\begin{centering}
\includegraphics[width=0.9\linewidth]{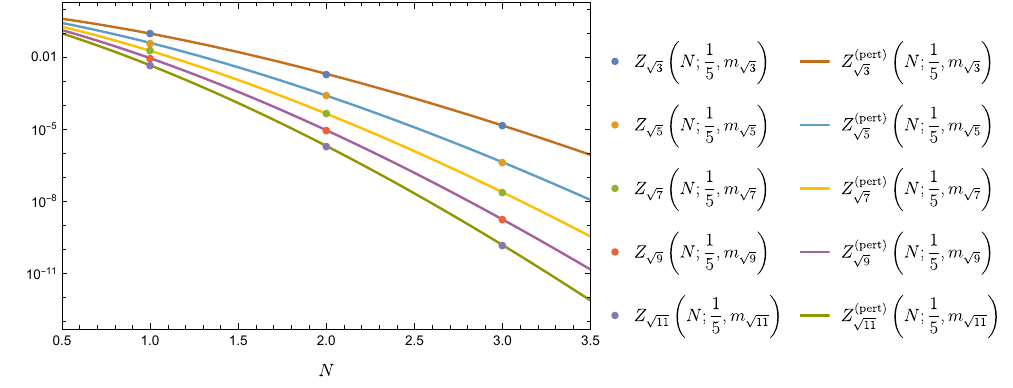}
\par\end{centering}
\caption{
Comparison between the values of the perturbative partition function \eqref{eq:zADHMpert} with \eqref{eq:ABC-FGF} (lines)
and the exact values of the partition function for $N=1,2,3$ \eqref{ZN1b2oddbyresiduesum}, \eqref{Zb2oddN2residuesum}, \eqref{Zb2oddN3residuesum} (dots)  at $b=\sqrt{3},\sqrt{5},\sqrt{7},\sqrt{9},\sqrt{11}$.
The FI and mass parameters are fixed to $\zeta=\frac{1}{5}$ and $m=m_b=\frac{b^2-3}{4b}i$.
The non-perturbative effects are suppressed even at $N=1$.
\label{fig:Pert-Exact-varyN}
}
\end{figure}
We perform this test by using the exact result for $N=1$ \eqref{eq:zADHM-N1}.
However, one may wonder $N=1$ is not large.
For answering this question, we first compare the perturbative partition function \eqref{eq:zADHMpert} with the exact values for $N=1,2,3$, $b=\sqrt{2n-1}$ and $m=m_b$ which are obtained from the finite sums \eqref{ZN1b2oddbyresiduesum}, \eqref{Zb2oddN2residuesum} and \eqref{Zb2oddN3residuesum}.
As displayed in Figure \ref{fig:Pert-Exact-varyN}, the perturbative partition function shows excellent agreement with the exact values for any value of $n$ and generic value of $\zeta$.
This suggests that our conjecture \eqref{eq:zADHMpert} with \eqref{eq:ABC-FGF} is indeed correct for $b=\sqrt{2n-1}$ and $m=m_b$.
Inversely, this suggests that the residue computation with the Young diagram is correct.
Notably, the agreement is good even at $N=1$ for which the partition function is obtained explicitly as \eqref{eq:zADHM-N1} for any positive real value of $b$.
This, however, would not be too surprising from the perspective of the instanton.
It is known that there are two types of the instantons, namely the worldsheet instanton and  the membrane instanton.
When $b=1$ and $m_1=m_2=m_3=0$, the correction of the worldsheet instanton is $\mathcal{O}\left( e^{-2\pi\sqrt{2N/k}} \right)$ \cite{Drukker:2010nc}, and the correction  of the membrane instanton is $\mathcal{O}\left( e^{-\pi\sqrt{2kN}} \right)$ \cite{Drukker:2011zy}.
These instanton effects have a clear holographic interpretation as the contributions from closed M2-branes wrapped on $S^7/\mathbb{Z}_k$, where the volume of the M2-branes precisely coincide with the instanton exponents.
We expect this interpretation persists with the squashing with $b>1$ and the mass deformations $m_1,m_2,m_3$, at least if these deformation parameters are not too large.
Namely, although these instanton effects would be
modified and also split into different species, the $k,N$-dependence of the exponents would remain the same.
In particular, for $k=1$, which is dual to the current setup, we expect that the instanton effects are sufficiently small even when $N$ is not large.
The similar phenomena were observed for the non-deformed ABJM theory with small $k$ \cite{Hatsuda:2012dt} or the ABJM theory with $k=1$ and $b^2=3$ \cite{Hatsuda:2016uqa}.
Note that in Figure \ref{fig:Pert-Exact-varyN} we have chosen $\zeta =1/5$ in order to stay away from the simplest case $\zeta =0$.

\begin{figure}
\begin{centering}
\includegraphics[width=0.47\linewidth]{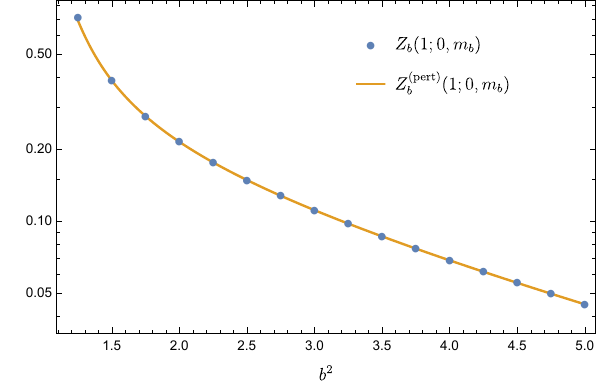} \includegraphics[width=0.47\linewidth]{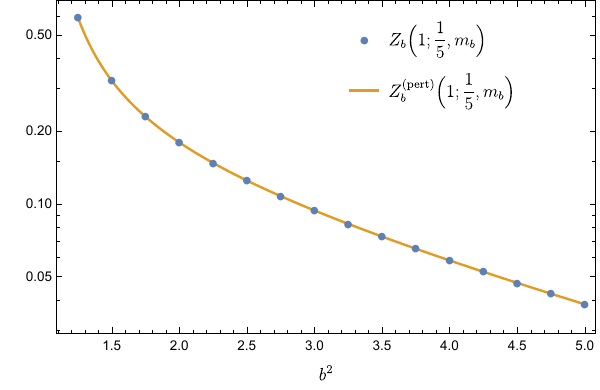}
\par\end{centering}
\caption{
The perturbative partition function \eqref{eq:zADHMpert} with \eqref{eq:ABC-FGF} vs the exact value for $N=1$ \eqref{eq:zADHM-N1}.
The FI and mass parameters are fixed to $\zeta=0,\frac{1}{5}$ (left/right) and $m=m_b=\frac{b^2-3}{4b}i$.
The perturbative partition function agrees with the $N=1$ exact value even for $b^2\notin 2\mathbb{N}-1$.
\label{fig:Pert-Exact-Genb}
}
\end{figure}
Encouraged by this observation, we use exact value at $N=1$ to check $b^2\notin 2\mathbb{N}-1$ case.
Namely, we compare the perturbative partition function \eqref{eq:zADHMpert} with the exact $N=1$ value \eqref{eq:zADHM-N1}, see Figure \ref{fig:Pert-Exact-Genb}.
Note that we avoid around $b\sim 1$ since it diverges as commented in section \ref{subsecLargeN}.
Note also that when we evaluate \eqref{eq:zADHM-N1}, we need to get the value of the $\mathcal{D}_{b}$ function.
Although this function is not simplified for general $b$, it can be computed numerically by using \eqref{eq:D_Int} or \eqref{eq:D_sum}.
As expected, we find that the perturbative partition function \eqref{eq:zADHMpert} agrees with the exact value at $N=1$ even for $b^2\notin 2\mathbb{N}-1$.
This result strongly suggests that \eqref{eq:zADHMpert} with \eqref{eq:ABC-FGF} gives the correct large $N$ expansion even for $b>1$.

\begin{figure}[t]
\begin{centering}
\includegraphics[width=0.47\linewidth]{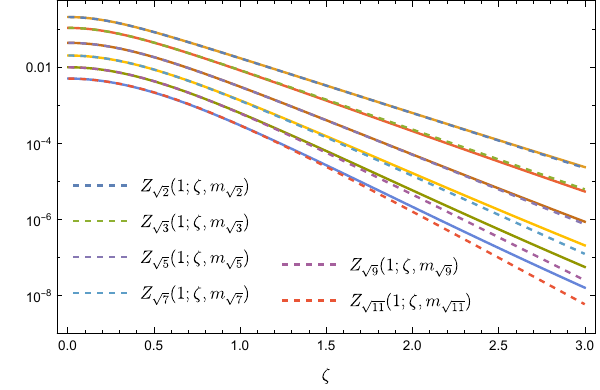}
\includegraphics[width=0.47\linewidth]{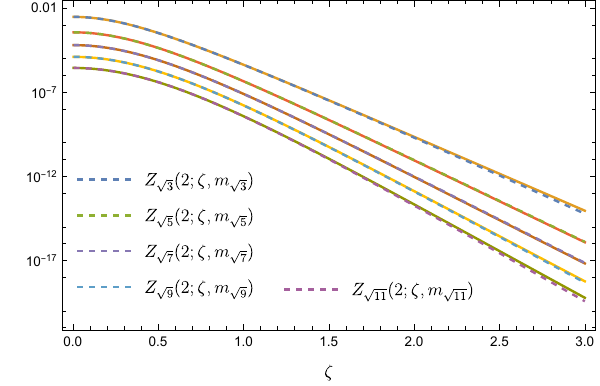} \par
\includegraphics[width=0.47\linewidth]{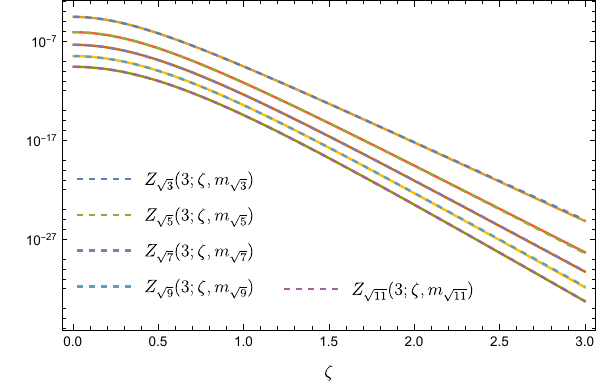}
\par
\end{centering}
\caption{
Comparison between the exact partition function \eqref{ZN1b2oddbyresiduesum}, \eqref{Zb2oddN2residuesum}, \eqref{Zb2oddN3residuesum} (or \eqref{eq:zADHM-N1} for $b=\sqrt{2}$) (dashed lines) and the perturbative partition function \eqref{eq:zADHMpert} (solid line accompanying with each dashed line) for varying FI parameter $\zeta$.
They are compatible with each other.
\label{fig:deviationfromPert-varyzetapart1}
}
\end{figure}
Next, since we have obtained the perturbative partition function \eqref{eq:zADHMpert} as a function of not only the squashing parameter but also the FI parameter, we compare the perturbative partition function and the exact partition functions obtained in section \ref{sec:NumericComp} as a function of the FI parameter $\zeta$.
The exact partition functions are obtained for odd $b^2$ with $N=1,2,3$ as \eqref{ZN1b2oddbyresiduesum}, \eqref{Zb2oddN2residuesum}, \eqref{Zb2oddN3residuesum}.
We also consider $b=\sqrt{2}$ by using \eqref{eq:zADHM-N1} since for $b=\sqrt{2}$ the $\mathcal{D}_b$ function can be written more explicitly than the integration/expansion formulas \eqref{eq:D_Int}/\eqref{eq:D_sum} as \eqref{eq:calDb2}.
Note that the value of $\mathcal{D}_{\sqrt{2}}(m)$ at $m=m_{\sqrt{2}}=-\frac{i}{4\sqrt{2}}$ \eqref{eq:Mass-FGF}, which appears in $Z_{\sqrt{2}}(1;\zeta,m_{\sqrt{2}})$, is obtained by taking an appropriate limit as
\begin{equation}
\mathcal{D}_{\sqrt{2}}\left(-\frac{i}{4\sqrt{2}}\right)=\frac{1}{2}.
\end{equation}
See Figure \ref{fig:deviationfromPert-varyzetapart1}.
We again find that the perturbative partition function matches the exact partition functions for $N=1,2,3$, which is a cross-check of our two results obtained in sections \ref{sec:LargeN} and \ref{sec:NumericComp}.

Lastly, let us also compare the exact values of the partition functions with the Airy form \eqref{eq:zABJMpert} for $b=\sqrt{2n-1}$ and generic real values of $m$.
In these cases the parameters in the Airy form are conjectured only up to the overall factor $e^{A_{1,b}(\zeta-m,\zeta+m,0)}$ as \eqref{eq:BC-ADHM}.
Nevertheless, it is possible to compare the ratio of the exact partition function $\frac{Z_b(N_1;\zeta,m)}{Z_b(N_2;\zeta,m)}$ at two different ranks $N_1,N_2$ with the ratio of the perturbative partition function
\begin{align}
\frac{Z_b^{\text{(pert)}}(N_1;\zeta,m)}{Z_b^{\text{(pert)}}(N_2;\zeta,m)}=\frac{\text{Ai}[C_{1,b}(\zeta-m,\zeta+m,0)^{-\frac{1}{3}}(N_1-B_{1,b}(\zeta-m,\zeta+m,0))]}{\text{Ai}[C_{1,b}(\zeta-m,\zeta+m,0)^{-\frac{1}{3}}(N_2-B_{1,b}(\zeta-m,\zeta+m,0))]},
\end{align}
with $B_{1,b},C_{1,b}$ in \eqref{eq:BC-ADHM}.
Here we choose $N_2=3$ and display the deviation of the ratio from 1
\begin{align}
\frac{
Z_b\left(N_1;m,\zeta\right)
}{
Z_b\left(3;m,\zeta\right)
}
\frac{
Z_b^{\text{(pert)}}\left(3;m,\zeta\right)
}{
Z_b^{\text{(pert)}}\left(N_1;m,\zeta\right)
}
-1,\label{eq:deviation}
\end{align}
for $m=\zeta=\frac{3}{10}$ and $m=\zeta=\frac{1}{2}$
\begin{align}
\begin{tabular}
{|c||c|c|c|c|}
\hline
            &$b=\sqrt{3}$&$b=\sqrt{5}$&$b=\sqrt{7}$&$b=\sqrt{9}$\\ \hline\hline
$m=\zeta=\frac{1}{5},N_1=1$ &$-2.532\times 10^{-5}$&  $-3.270\times 10^{-5}$  &$-4.672\times 10^{-5}$&$-6.757\times 10^{-5}$\\ \hline
$m=\zeta=\frac{1}{5},N_1=2$ &$1.899\times 10^{-6}$ &  $2.930\times 10^{-6}$   &$3.035\times 10^{-6}$ &$2.510\times 10^{-6}$\\ \hline\hline
$m=\zeta=\frac{3}{10},N_1=1$&$8.576\times 10^{-4}$ &  $8.043\times 10^{-4}$   &$7.242\times 10^{-4}$ &$6.471\times 10^{-4}$\\ \hline
$m=\zeta=\frac{3}{10},N_1=2$&$-2.991\times 10^{-5}$&  $-1.022\times 10^{-6}$ &$1.583\times 10^{-5}$ &$2.534\times 10^{-5}$\\ \hline
\end{tabular}
\end{align}
When $m$ and $\zeta$ are sufficiently small, we find good agreement between exact value and Airy form, which also improves as $N_1$ increases.
This suggests that the Airy form gives the correct large $N$ expansion when $m$ and $\zeta$ are small.

\begin{figure}[t]
\begin{centering}
\includegraphics[width=0.47\linewidth]
{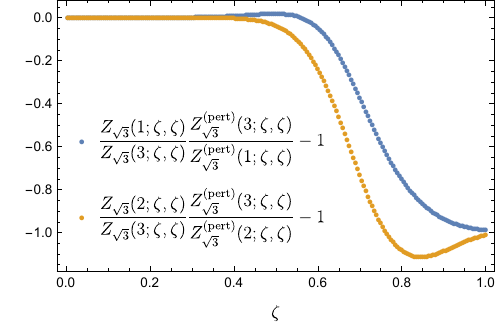}
\includegraphics[width=0.47\linewidth]
{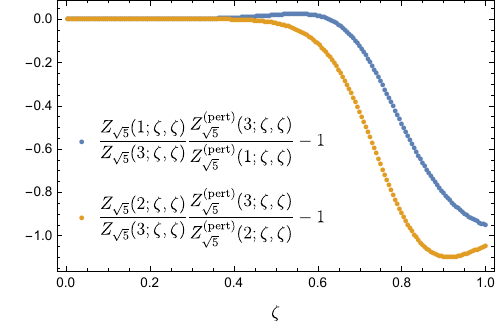} \par
\includegraphics[width=0.47\linewidth]
{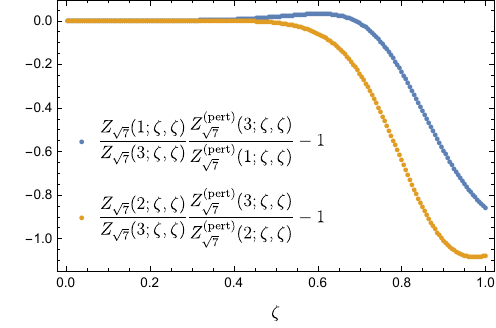}
\includegraphics[width=0.47\linewidth]
{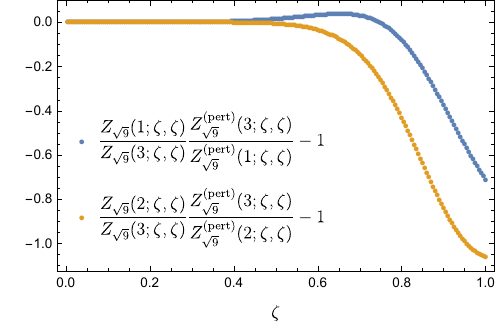}
\par\end{centering}
\caption{
The deviation of the ratio between $\frac{Z_b(N;\zeta,m)}{Z_b(3;\zeta,m)}$ and $\frac{Z_b^{\text{(pert)}}(N;\zeta,m)}{Z_b^{\text{(pert)}}(3;\zeta,m)}$ from $1$ \eqref{eq:deviation} for $b=\sqrt{2n-1}$ with $n\in\mathbb{Z}$ and $\zeta=m$.
We find that the deviation is getting significant around $\zeta=\frac{1}{2}$.
}
\label{bothmandFI_vsAirypart1}
\end{figure}
On the other hand, as $m=\zeta$ increases we observe that the deviation becomes significant around $m=\zeta=\frac{1}{2}$, as displayed in Figure \ref{bothmandFI_vsAirypart1}.
Indeed, the same deviation was also found for $b=1$, where it was conjectured that the partition function exhibits a large $N$ phase transition at $m\zeta=\frac{1}{4}$ \cite{Nosaka:2016vqf,Honda:2018pqa}.
From the large $N$ expansion in the sub-critical regime, this phase transition is also viewed as the instability of the non-perturbative effects corresponding to M2-branes wrapped on $S^7/\mathbb{Z}_k$ \cite{Nosaka:2015iiw}.
It would be interesting to investigate the phase structure and the $1/N$ non-perturbative effects in the mass deformed ABJM theory also on the squashed sphere.\footnote{
See \cite{Shimizu:2018evl} for an evidence of large $N$ phase transition on squashed sphere through the large $N$ saddle point approximation.
}

\section{Conclusion\label{sec:Conclusion}}

In this paper we have analytically obtained the exact large $N$ expansion of the partition function of the mass deformed ABJM theory on the squashed three sphere, which still takes the form of the Airy function.
Although the Chern-Simons level and two out of three mass parameters are fixed in our study, we could still retain the squashing parameter $b$ and the remaining mass parameter corresponding to the FI parameter $\zeta$ in the dual $\mathcal{N}=4$ SYM as free parameters.
We have carried out the computation by applying the Fermi gas formalism to the dual SYM matrix model with odd $b^{2}$ and observed that the numerical comparison strongly indicated that our main result \eqref{eq:zADHMpert} with \eqref{eq:ABC-FGF} should hold for arbitrary $b>1$.
We have also proposed a systematic way of computing the exact partition function of SYM matrix model on $S^3_b$, including the non-perturbative contributions, up to $N=3$, where the residue contributions are classified in terms of Young diagrams.
This exact result matches with the aforementioned perturbative expansion in the regime where non-perturbative contributions are suppressed, providing a strong evidence that both the Fermi gas approach and the Young diagram approach work well.

There are various interesting directions for further study.
First, although in this paper we have focused on the perturbative part, in general the Fermi gas formalism can also be used for investigating the non-perturbative effect (see for example the original paper of the Fermi gas formalism \cite{Marino:2011eh}).
In type IIA string theory, they are identified with a fundamental string wrapping the $\mathbb{CP}^1$ inside $\mathbb{CP}^3$ or a D2-brane
wrapping the $\mathbb{RP}^3$ inside $\mathbb{CP}^3$ \cite{Cagnazzo:2009zh,Drukker:2010nc,Drukker:2011zy}.
For the round sphere case, the non-perturbative effects corresponding to the fundamental strings have been studied in type IIA string theory or M-theory including the instanton coefficients, and the result (partially) matches to the matrix model computation \cite{Gautason:2023igo,Beccaria:2023ujc}.
It would be interesting if one can perform the same check for the squashed case.

Second, it would be nice if one can generalize our result.
For example, since we had to fix the parameters of the ABJM theory $(k,m,m_3)$
(and the two ranks $(N_1,N_2)$ of the gauge group $\mathrm{U}(N_1) \times \mathrm{U}(N_2)$ to be equal)
in our approach, finding exact perturbative results for completely general parameters is still a remaining task.
As other generalizations, because various theories related to the M2-branes are known, it is interesting whether the Fermi gas approach is applicable also for these theories.
For example, one can consider the case when the number of the matters of the $\mathcal{N}=4$ SYM theory is greater than one.
The squashing dependence of the large $N$ behavior of this matrix model has been actually conjectured in \cite{Bobev:2023lkx}, and thus one probably can reproduce the conjecture rigorously along our approach.
One can also consider the case when each node of the gauge group can be other than the unitary group \cite{Aharony:2008gk,Hosomichi:2008jb} or the form of the quiver diagram can be generalized to affine ADE Dynkin diagrams \cite{Gulotta:2011vp}.
Although the large $N$ behavior of the matrix model of these theories with $b=1$ has been well studied, the one with $b\neq 1$ has not been investigated with the first-principle calculation in literature.

Third, in \cite{Bonelli:2016idi,Bonelli:2017gdk,
Nosaka:2020tyv,Nosaka:2024gle} it was found that the partition function of $\text{U}(N)_k\times \text{U}(N+M)_{-k}$ ABJM theory with two-parameter mass deformation \eqref{eq:zABJM-m30} on round sphere satisfies non-linear relations among different ranks $N,M$.
Written in terms of the grand canonical partition function with respect to $N$, these relations are summarized to a bilinear difference relation with respect to $M$ which is similar to the $q$-discrete Toda equation in $\tau$-form.
In particular, combining with duality cascade \cite{Evslin:2009pk,Aharony:2009fc,Honda:2020uou}, we obtain a non-trivial bilinear relation for the grand canonical partition function of the $\text{U}(N)_1\times \text{U}(N)_{-1}$ ABJM theory which is dual to the $\mathcal{N}=4$ SYM.
It would be interesting to investigate such relation also for $\mathcal{N}=4$ SYM on squashed sphere.
The closed form expressions of the partition function for $b^2\in 2\mathbb{N}-1$ obtained in section \ref{sec:NumericComp} would be helpful for this purpose.

\section*{Acknowledgements}

We are grateful to Kiril Hristov and Tadashi Okazaki for valuable discussions.
NK and YP are supported by National key research and development program under grant No.~2022YFE0134300 and the National Natural Science Foundation of China
(NSFC) under grants No.~12175164 and No.~12247103.
The work of T.N.~was supported by the Startup Funding no.~2302-SRFP-2024-0012 of Shanghai Institute for Mathematics and Interdisciplinary Sciences.

\appendix

\section{Double sine function and related function\label{sec:DSF}}

In this section we summarize properties of the double sine function and a related function defined by a ratio of two double sine functions \cite{Bytsko:2006ut}.

The double sine function is defined as
\begin{equation}
s_{b}\left(z\right)=\prod_{m,n=0}^{\infty}\frac{mb+nb^{-1}+\frac{Q}{2}-iz}{mb+nb^{-1}+\frac{Q}{2}+iz},\quad Q=b+b^{-1}.
\end{equation}
The double sine function also has an integral expression
\begin{equation}
s_{b}\left(z\right)=\exp\left[-\frac{i\pi}{2}z^{2}-\frac{i\pi}{24}\left(b^{2}+b^{-2}\right)+\int_{\mathbb{R}+i\epsilon}\frac{dt}{t}\frac{e^{-2izt}}{4\sinh\left(bt\right)\sinh\left(b^{-1}t\right)}\right].\label{eq:DSF-Def2}
\end{equation}
The double sine function satisfies the following identities
\begin{equation}
s_{b}\left(0\right)=1,\quad s_{b}\left(z\right)=s_{b^{-1}}\left(z\right),\quad s_{b}\left(z\right)s_{b}\left(-z\right)=1,\quad\overline{s_{b}\left(z\right)}=s_{b}\left(-\bar{z}\right).\label{eq:DSF-props}
\end{equation}
The most important property in this paper is
\begin{equation}
\frac{s_{b}\left(z+\frac{i}{2}b^{\pm1}\right)}{s_{b}\left(z-\frac{i}{2}b^{\pm1}\right)}=\frac{1}{2\cosh\left(\pi b^{\pm1}z\right)}.\label{eq:DSF-Cosh}
\end{equation}

We also define a $\mathcal{D}_b$ function \cite{Bytsko:2006ut,Hatsuda:2016uqa}
\begin{equation}
\mathcal{D}_{b}\left(z\right)=\frac{s_{b}\left(z+\frac{i}{4}Q\right)}{s_{b}\left(z-\frac{i}{4}Q\right)}.\label{eq:D-Def}
\end{equation}
This function is parity even
\begin{equation}
\mathcal{D}_{b}\left(z\right)=\mathcal{D}_{b}\left(-z\right),\label{eq:D-parity}
\end{equation}
and invariant under the Fourier transformation
\begin{equation}
\int_\mathbb{R}dxe^{2\pi ixy}\mathcal{D}_{b}\left(x\right)=\mathcal{D}_{b}\left(y\right).\label{eq:D-Fourier}
\end{equation}
When $b=1$, $\mathcal{D}_b$ is simplified by using \eqref{eq:DSF-Cosh}. When $b^{2}$ is positive odd integer, say $b^{2}=2n-1$, we can also simplify $\mathcal{D}_b$ by using \eqref{eq:DSF-Cosh} repeatedly as
\begin{equation}
\mathcal{D}_{b=\sqrt{2n-1}}\left(\mu\right)=\prod_{\ell=1}^{n}\frac{1}{2\cosh\left(\frac{\pi}{b}\mu+\frac{\pi i}{b^{2}}\left(\frac{n+1}{2}-\ell\right)\right)}.\label{eq:D-bOdd}
\end{equation}
Another special value where $\mathcal{D}_b$ simplifies is $b^2=2$ \cite{Hatsuda:2016uqa}
\begin{align}
\mathcal{D}_{\sqrt{2}}\left(\mu\right) & =\frac{1}{2^{\frac{1}{4}}\left(2\cosh\left(2\sqrt{2}\pi\mu\right)\right)^{\frac{1}{8}}\left(\sqrt{2}\cosh\left(\sqrt{2}\pi\mu\right)+1\right)^{\frac{1}{2}}}\nonumber \\
 & \quad\times\exp\left[-\sqrt{2}\mu\arctan\left(e^{-2\sqrt{2}\pi\mu}\right)+\frac{i}{4\pi}\left(\mathrm{Li}_{2}\left(ie^{-2\sqrt{2}\pi\mu}\right)-\mathrm{Li}_{2}\left(-ie^{-2\sqrt{2}\pi\mu}\right)\right)\right].\label{eq:calDb2}
\end{align}

For generic $b$, the $\mathcal{D}_b$ function can be evaluated numerically by the integral expression
\begin{equation}
\mathcal{D}_{b}\left(z\right)=\exp\left(\int_{\mathbb{R}+i\epsilon}\frac{dt}{t}\frac{\sinh\frac{Qt}{2}\cos\left(2zt\right)}{2\sinh\left(bt\right)\sinh\left(b^{-1}t\right)}\right),\label{eq:D_Int}
\end{equation}
which is obtained by combining \eqref{eq:DSF-Def2} with \eqref{eq:D-parity}.
For $\text{Re}\left[z\right]\neq 0$, we can also evaluate $\mathcal{D}_b\left(z\right)$ by the following formula
\begin{align}
\mathcal{D}_b\left(z\right)=\text{exp}\left[
-\frac{\pi Q\left(\pm z\right)}{2}
+\sum_{\ell=1}^\infty\left(\frac{e^{-2\pi \ell  b\left(\pm z\right)}}{2\ell\cos\frac{\pi \ell bQ}{2}}
+\frac{e^{-\frac{2\pi \ell \left(\pm z\right)}{b}}}{2\ell\cos\frac{\pi \ell Q}{2b}}
\right)
\right],
\label{eq:D_sum}
\end{align}
where the sign $\pm$ is chosen such that $\text{Re}\left[\pm z\right]>0$.

\section{Duality between ABJM and $\mathcal{N}=4$ SYM theories\label{sec:Duality}}

In this section we consider the duality between the ABJM theory with $k=1$ and the $\mathcal{N}=4$ $\mathrm{U}\left(N\right)$ SYM theory with an adjoint hypermultiplet and a fundamental hypermultiplet and show the parameter correspondence \eqref{eq:Duality-Para}. We consider two cases, namely, when $b=1$ or $N=1$.

First, we consider the case when $b=1$ \cite{Kapustin:2010xq}.
In this case, the $\mathcal{D}_b$ function is simplified thanks to \eqref{eq:D-bOdd}.
The strategy is to apply the Fermi gas formalism also to the ABJM theory \cite{Nosaka:2015iiw}.
The ABJM matrix model \eqref{eq:zABJM-m30} with $k=1$ is
\begin{align}
&Z_{1,1}^{\mathrm{ABJM}}\left(N;m_{1},m_{2},0\right) \nonumber \\
& =\frac{1}{\left(N!\right)^{2}}\int_{\mathbb{R}}\prod_{i}^{N}\frac{d\mu_{i}}{2\pi}\frac{d\nu_{i}}{2\pi}e^{-\frac{i}{4\pi}\sum_{i}\left(\mu_{i}^{2}-\nu_{i}^{2}\right)}\nonumber \\
 & \quad\quad\times\frac{\prod_{i<j}^{N}2\sinh\frac{\mu_{i}-\mu_{j}}{2}\prod_{i<j}^{N}2\sinh\frac{\nu_{i}-\nu_{j}}{2}}{\prod_{i,j}^{N}2\cosh\frac{\mu_{i}-\nu_{j}+\pi\left(m_{2}+m_{1}\right)}{2}}\frac{\prod_{i<j}^{N}2\sinh\frac{\nu_{i}-\nu_{j}}{2}\prod_{i<j}^{N}2\sinh\frac{\mu_{i}-\mu_{j}}{2}}{\prod_{i,j}^{N}2\cosh\frac{\nu_{i}-\mu_{j}+\pi\left(m_{2}-m_{1}\right)}{2}}.\label{eq:MM-ABJM-Def}
\end{align}
By using \eqref{eq:CauchyDetForm} and \eqref{eq:Cosh-Op}, one obtains
\begin{align}
Z_{1,1}^{\mathrm{ABJM}}\left(N;m_{1},m_{2},0\right) & =\frac{1}{\left(N!\right)^{2}}\int_{\mathbb{R}}\prod_{i}^{N}d\mu_{i}\det\left(\left[\braket{\mu_{i}|e^{-\frac{i}{4\pi}\hat{x}^{2}}\frac{e^{i\frac{m_{2}+m_{1}}{2}\hat{p}}}{2\cosh\frac{\hat{p}}{2}}e^{\frac{i}{4\pi}\hat{x}^{2}}|\nu_{j}}\right]_{i,j}^{N\times N}\right)\nonumber \\
 & \quad\quad\times\det\left(\left[\braket{\nu_{i}|\frac{e^{i\frac{m_{2}-m_{1}}{2}\hat{p}}}{2\cosh\frac{\hat{p}}{2}}|\mu_{j}}\right]_{i,j}^{N\times N}\right),
\end{align}
where $\left[\hat{x},\hat{p}\right]=2\pi i$.
By using a gluing formula
\begin{align}
&\frac{1}{N!}\int_\mathbb{R}\prod_{i}^{N}d\beta_{i}\det\left(\left[\braket{\alpha_{i}|\hat{A}|\beta_{j}}\right]_{i,j}^{N\times N}\right)\det\left(\left[\braket{\beta_{i}|\hat{B}|\gamma_{j}}\right]_{i,j}^{N\times N}\right)\nonumber \\
&=\det\left(\left[\braket{\alpha_{i}|\hat{A}\hat{B}|\gamma_{j}}\right]_{i,j}^{N\times N}\right),
\end{align}
we obtain
\begin{equation}
Z_{1,1}^{\mathrm{ABJM}}\left(N;m_{1},m_{2},0\right)=\frac{1}{N!}\int_{\mathbb{R}}\prod_{i}^{N}d\mu_{i}\det\left(\left[\braket{\mu_{i}|e^{-\frac{i}{4\pi}\hat{x}^{2}}\frac{e^{i\frac{m_{2}+m_{1}}{2}\hat{p}}}{2\cosh\frac{\hat{p}}{2}}e^{\frac{i}{4\pi}\hat{x}^{2}}\frac{e^{i\frac{m_{2}-m_{1}}{2}\hat{p}}}{2\cosh\frac{\hat{p}}{2}}|\mu_{j}}\right]_{i,j}^{N\times N}\right).
\end{equation}
After taking a similarity transformation by $e^{\frac{i}{4\pi}\hat{p}^{2}}$, by using
\begin{equation}
e^{-\frac{i}{4\pi}\hat{p}^{2}}e^{-\frac{i}{4\pi}\hat{x}^{2}}f\left(\hat{p}\right)e^{\frac{i}{4\pi}\hat{x}^{2}}e^{\frac{i}{4\pi}\hat{p}^{2}}=f\left(\hat{x}\right),
\end{equation}
we obtain the form of \eqref{eq:MM-FGS} with the density matrix
\begin{equation}
\hat{\rho}_{1,1}^{\mathrm{ABJM}}\left(m_{1},m_{2},m_3=0\right)=\frac{e^{i\frac{m_{2}+m_{1}}{2}\hat{x}}}{2\cosh\frac{\hat{x}}{2}}\frac{e^{i\frac{m_{2}-m_{1}}{2}\hat{p}}}{2\cosh\frac{\hat{p}}{2}}.
\end{equation}
On the other hand, the density matrix of the $\mathcal{N}=4$ SYM theory is \eqref{eq:DM-b1}.
Therefore, the parameter correspondence for $b=1$ is consistent with \eqref{eq:Duality-Para}.

Next, we consider the case when $N=1$.
The ABJM matrix model \eqref{eq:zABJM-m30} with $k=1$ is
\begin{equation}
Z_{1,b}^{\mathrm{ABJM}}\left(1;m_{1},m_{2},0\right)=\int_{\mathbb{R}}\frac{d\mu}{2\pi}\frac{d\nu}{2\pi}e^{-\frac{i}{4\pi}\left(\mu^{2}-\nu^{2}\right)}\mathcal{D}_{b}\left(\frac{\mu-\nu}{2\pi}+\frac{m_{2}+m_{1}}{2}\right)\mathcal{D}_{b}\left(\frac{\nu-\mu}{2\pi}+\frac{m_{2}-m_{1}}{2}\right).
\end{equation}
After shifting the integration variable as $\nu\rightarrow\nu+\mu$, one can perform the integration over $\mu$, which produce the delta function $\delta\left(\nu\right)$ via $\int dxe^{2\pi iyx}=\delta\left(y\right)$.
The delta function can be used for performing the integration over $\nu$, and thus we finally obtain
\begin{equation}
Z_{1,b}^{\mathrm{ABJM}}\left(1;m_{1},m_{2},0\right)=\mathcal{D}_{b}\left(\frac{m_{2}+m_{1}}{2}\right)\mathcal{D}_{b}\left(\frac{m_{2}-m_{1}}{2}\right).
\end{equation}
On the other hand, the SYM matrix model is \eqref{eq:zADHM-N1}. Therefore, when $N=1$, the parameter correspondence is again consistent with \eqref{eq:Duality-Para}.

\printbibliography

\end{document}